%% file: main_pdf.tex
\documentclass{amsart}
\UseRawInputEncoding
\usepackage{graphicx,xcolor}
\usepackage{adjustbox}
\usepackage{bm, amsmath}
\usepackage[authoryear]{natbib}
\usepackage[backref]{hyperref}
\usepackage{cleveref}
\usepackage{amsfonts}

\newtheorem{theorem}{Theorem}

\renewcommand{\vec}[1]{{\bm{#1}}}

\title[TFDA for the lid-cavity flow]{Topological flow data analysis for transient flow patterns: a graph-based approach}
\author[T. Sakajo]{Takashi Sakajo}
\address{Department of Mathematics, Kyoto University, Kyoto, 606-8502, Japan}
\email{sakajo@math.kyoto-u.ac.jp}

\author[T. Matsumoto]{Takeshi Matsumoto}
\address{Department of Physics, Kyoto University, Kyoto, 606-8502, Japan}

\author[S. Kaji]{Shizuo Kaji}
\address{Institute of Mathematics for Industry, Kyushu University, Fukuoka, 819-0395, Japan}

\author[T. Yokoyama]{Tomoo Yokoyama}
\address{Department of Mathematics, Saitama University, Saitama, 338-8570, Japan}

\author[T. Uda]{Tomoki Uda}
\address{Academic Assembly, University of Toyama, Toyama, 930-8555, Japan}

\date{}

\begin{document}

\maketitle

\begin{abstract}
We introduce a method of time series analysis for two-dimensional transient flow patterns based on Topological Flow Data Analysis (TFDA), a new approach to topological data analysis. 
TFDA identifies local topological flow structures from an instantaneous streamline pattern and describes their global connections as a unique planar tree and its string representation.
With TFDA, the evolution of two-dimensional flow patterns is reduced to a discrete dynamical system represented as a transition graph between topologically equivalent streamline patterns.
We apply this method to study the lid-driven cavity flow for Reynolds numbers from $Re=14000$ to $16000$, a benchmark problem in the analysis of fluid dynamics. 
{Our approach can extract some physical information from the lid-driven cavity flow: transition of the flow from periodic to quasi-periodic and chaotic; estimation of the period of periodic dynamics; relation between variations in energy and enstrophy and topological changes in flow patterns; statistical properties of intricate flow evolution at higher Reynolds number.
In addition, we perform an observational causal inference to analyse changes in local flow patterns in the cavity corner.}
This work demonstrates the potential of TFDA-based time series analysis to uncover complex dynamical behaviours in fluid flow data from a topological perspective.
\end{abstract}


\section{Introduction}\label{sec:1}
 Investigations of fluid dynamics are not complete when we finish collecting data on the flow field, such as velocity, pressure, vorticity, and so forth.
 To describe and understand the rich behaviours of a flow in a given condition, we need to find a structure in the data.
 Some structures are easily caught by our eyes, such as a vortex sheet or tube, which can be highlighted by certain visualization methods in laboratory experiments or numerical simulations. 
 A hairpin vortex in the boundary layer turbulence is a primary example of those structures that lead to the current understanding of turbulence, see, e.g., \cite{da15}.   
 However, in general, finding a relevant structure for a given flow is difficult. 
 Moreover, finding a reduced-order model of the flow with respect to the elements of the structure is even harder. 
 For reduced-order modelling, the Galerkin projection with a handful of terms of the known governing equation has been a systematic tool. 
 Data-driven approaches without needing to know the governing equation include the so-called proper orthogonal decomposition (POD) and dynamic mode decomposition (DMD), as explained in \cite{bk22} for instance. 

Topological Data Analysis (TDA) is a relatively new method that describes the topological features of data. 
See an introductory book by~\cite{edelsbrunner2010computational} and an advanced book by~\cite{Dey_Wang_2022}. 
While TDA applications in fluid dynamics have gained attention in recent years, concrete examples of their implementation remain limited. 
In the present paper, we construct a reduced model of the temporal evolution of streamline patterns using a different TDA approach, called Topological Flow Data Analysis (TFDA), which focuses on the topology of instantaneous streamline patterns of two-dimensional flows.
Through this model, we quantitatively track the temporal changes in the topological features of the flow field and analyse the causal relationships between local flow structures by treating pattern transitions as a discrete dynamical system on the reduced model. 
We discuss both the effectiveness of this method and its potential for future development.

The mathematical background behind the TFDA is explained as follows. 
Let $\psi({\bm x})$ be a streamfunction of $C^r$-class ($r\ge 1$) on a two-dimensional domain $\mathcal{D} \subset \mathbb{R}^2$. 
Since the vector field is defined by ${\bm u}=-\nabla^\perp \psi$ with $\nabla^\perp = (\partial_y, -\partial_x)$, it is considered a Hamiltonian vector field with $\psi$ being its Hamiltonian. 
Let $O({\bm x}_0)$ denote a particle orbit advected by the vector field ${\bm u}({\bm x})$ starting from ${\bm x}_0 \in \mathcal{D}$.
 That is, $O({\bm x}_0)=\{ {\bm \phi}(s; {\bm x}_0) \, \vert \, s \in \mathbb{R} \}$, where ${\bm \phi}(s; {\bm x}_0)$ is the unique solution to the initial value problem $\dot{\bm \phi}(s) = {\bm u}( {\bm \phi}(s))$ and ${\bm \phi}(0)={\bm x}_0$. 
 When the vector field is stationary, the particle orbit is contained in a level curve of the streamfunction $\psi$, i.e., streamlines, owing to $\nabla \psi \cdot {\bm u}=0$.
 When the velocity field depends on time, particle orbits do not correspond to streamlines in general. 
 However, we consider the topological structure of instantaneous streamlines every fixed time.
A mathematical theory for classifying topological features of the set of all instantaneous streamlines,
 $\mathcal{O}=\{ O({\bm x}_0) \, \vert \, {\bm x}_0 \in \mathcal{D}\}$, has been developed by \cite{TFDA-YS12, TFDA-SY18}.
 The topological classification is realised under the restrictions of Poincar\'e--Bendixson theorem, which gives a classification of fixed points, periodic orbits, and limit orbits as $s \rightarrow \pm \infty$, and Poincar\'e--Hopf theorem, which defines the relationship between the index of the vector field and the Euler characteristic of the flow domain. 
 Based on the theory, it is shown in~\cite{TFDA-UYS18} that the topological structure of instantaneous streamlines in $\mathcal{O}$ is uniquely converted into a discrete graph, named a partially Cyclically Ordered rooted Tree (the abbreviation {\it COT} is used in this paper), and its string expression, called {\it COT representation}. Through this transformation, the topological patterns of flow streamlines are represented as graphs or strings, which we call Topological Flow Data Analysis (TFDA).
 It can also describe the evolution of complex flow patterns as a discrete dynamical system of trees, which gives rise to a reduced-order model of complex flow evolution in terms of topology, as we will demonstrate in this paper.
 Note that TFDA was originally developed for Hamiltonian vector fields as above, but it has recently been extended to two-dimensional compressible vector fields~\cite{TFDA-SY22}.

TFDA has been applied to a variety of fluid-mechanical problems.
In \cite{ssy14}, TFDA was used to analyse boundary layer separations around a flat plate with the Reynolds number $1370$ based on chord length and angle of attack $15^\circ$ in a numerical simulation of a two-dimensional flow. 
The dynamical process was described with brief COT representations. 
In \cite{TFDA-USIK21}, TFDA was used to detect an atmospheric blocking phenomenon, a long-lived high-pressure region that blocks the westerly jet flow in the surrounding region, with observational data of a year on the synoptic scale of more than $1000$ km. 
It was confirmed that the TFDA method outperformed other existing methods for detecting blocking and that the morphological types of blocking were identified with the COT representations in accordance with the meteorologists' empirical classifications. 
\cite{TFDA-SOU22} considered Kuroshio Large Meander in oceanography, which is an abnormal current that causes the capture of a cool cyclonic eddy with substantial impacts on fisheries and marine transport.
TFDA is applied to sea surface height data to detect the occurrence of an abnormal state.
{Recently, \cite{Kimura2026} has applied a TFDA to two dimensional turbulent flows in a doubly periodic box, allowing us to track coherent vortex structures statistically.}
Furthermore, TFDA has been utilised more complicated flow patterns in the left ventricle of the human heart{~\cite{TFDA-SI23, Dyer2026}.}
Actual measurement data with echocardiography (two spatial dimensions and time) and magnetic resonance imaging (a slice of three spatial dimensions and time) were analysed. 
Specifically, a swirling flow domain is defined as a topological vortex structure using COT representations, and it was shown that differences in flow patterns between normal and diseased hearts were resolved objectively with topological vortex structures in agreement with the diagnoses of experienced cardiovascular surgeons.

In these applications mentioned above, TFDA was mainly used to state identifications of flow patterns with vortex-like flow structures, but methods for characterizing the dynamics of transient flow patterns using TFDA have not yet been fully developed.
Therefore, in this paper, we propose a new time series analysis based on TFDA to describe the complex dynamics exhibited by transient flow patterns. 
Our target flow is simple and prototypical: the two-dimensional lid-driven cavity flow~\cite{shen91}. 
It is perhaps the most famous benchmark problem in computational fluid mechanics when the flow is in a nonlinear steady state consisting of several corner eddies and one big central eddy. 
The flow is also considered a representative example of the complexity of the recirculating eddy and flow separation in a textbook \cite{rg13}. 
Those eddy structures are explicitly expressed in the COT representation, as we see later.
This means that the graph representation obtained via TFDA is always interpretable in contrast to, for example, the singular vectors of POD and DMD, which are not always easy to interpret.
In particular, we demonstrate the capability of TFDA by applying it to numerical lid-driven cavity flow with the transitional range of Reynolds numbers from time-periodic states to chaotic states.

 The paper is constructed as follows. 
 In Section~\ref{sec:2}, after reviewing the lid-driven cavity flow and the numerical methods used here, we provide how we apply TFDA to the numerical data. 
{Section~\ref{sec:3} provides a detailed time series analysis using TFDA when the Reynolds number is $Re=14000$.
We identify the change in topological flow patterns from the COT representations and show that flow dynamics is reduced to a transition graph between topologically equivalent flow patterns, called topological states.
In Section~\ref{sec:4}, we investigate the physics of the lid-driven cavity flow across Reynolds numbers from $Re=14000$ to $16000$, where the flow dynamics changes from periodic to quasi-periodic and chaotic.
We first show how the flow dynamics becomes complicated in terms of the number of topological states as the Reynolds number increases. 
Second, we estimate the period of periodic flow dynamics from the time series of COT representations.
Third, we relate the variations of physical quantities such as energy and enstrophy to the transitions of topological states.
Fourth, we assign serial numbers (IDs) to topological states in the order of their appearance across Reynolds numbers, and examine whether these states persist in the transition to chaotic flow through their occurrence rates.}
Finally, to take advantage of the high interpretability of TFDA, we perform an observational causal analysis based on the time series of COT representations.
We show that the observational approach based on TFDA detects a geometric causality between local topological patterns in chaotic flow that the standard sensitivity analysis cannot capture, as compared in Appendix \ref{sec:Appendix}.
The discussion and concluding remarks are in Section~\ref{sec:5}.

\section{Our target and method}\label{sec:2}
\subsection{The lid-driven cavity flow\label{subsec:2.1}}
\begin{figure}
    \centering
    \includegraphics[width=1.0\linewidth]{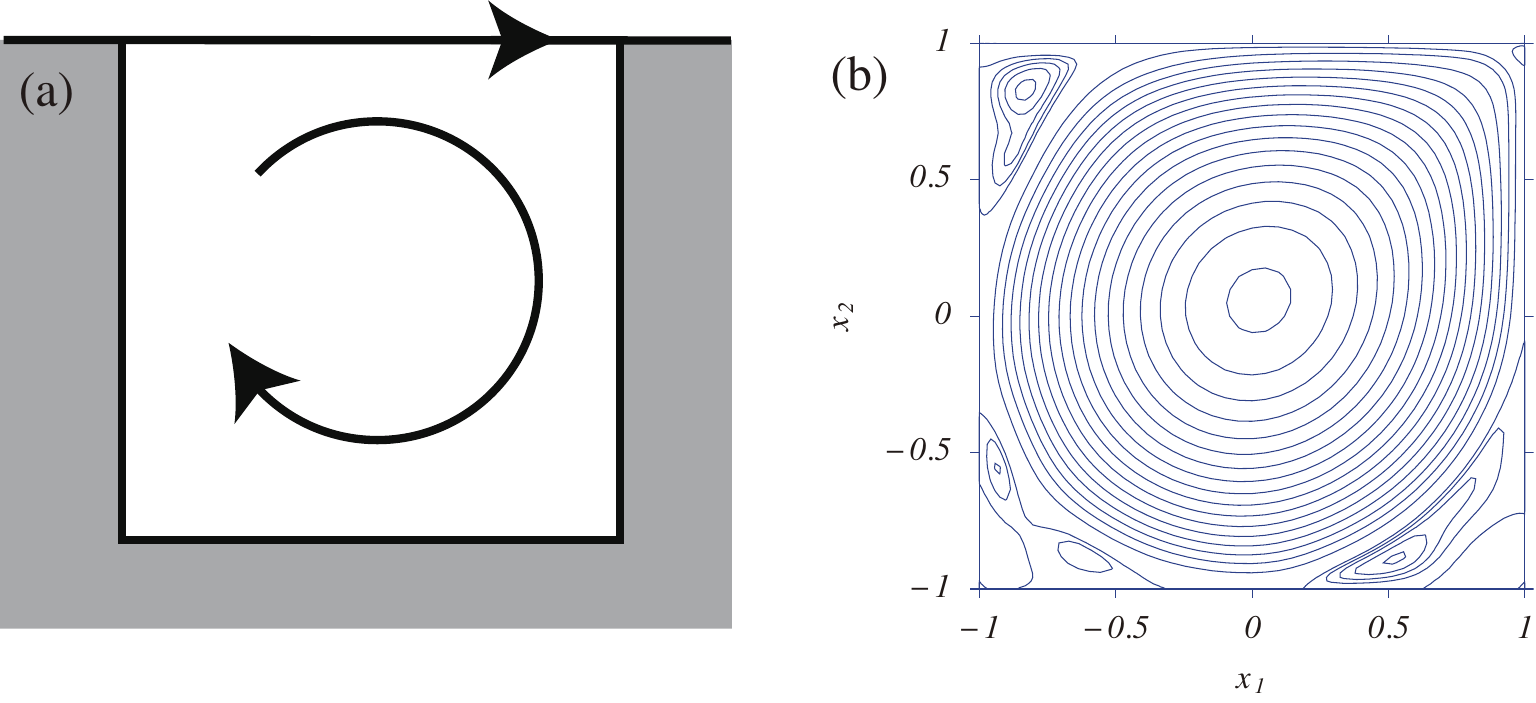}
    \caption{(a) The lid-driven cavity flow. (b) A snapshot of a streamline pattern in the cavity.}
    \label{fig:cavity}
\end{figure}

The lid-driven cavity flow is a viscous flow in a hollow space enclosed by no-slip walls. 
The shape of the cavity is usually set to a rectangle in two spatial dimensions and to a rectangular cuboid in three dimensions.
One of the walls, the lid, moves with a constant velocity and other walls remain still (zero velocity), as shown in Figure~\ref{fig:cavity}(a). 
As the energy injected constantly by the lid is dissipated by the viscosity,  the flow reaches a variety
of states depending on the Reynolds number as reviewed in \cite{sd00}. 
For low Reynolds numbers, the flows are stationary. In other words,
they are nonlinearly stationary solutions to the Navier-Stokes equations with the given boundary condition.
Those stationary solutions, in particular two-dimensional ones, serve as a standard benchmark for numerical methods; see, e.g., \cite{ghia82} and \cite{bp98}. As we increase the Reynolds number, the flow becomes time-periodic and then turbulent \cite{sd00}.

A fundamental fluid-dynamical characteristic of the lid-driven cavity flow, whether stationary, periodic, or turbulent, is the formation of recirculating eddies at the corners, as emphasized by \cite{rg13}. 
This phenomenon is illustrated in Figure~\ref{fig:cavity}(b), which shows the streamfunction of a stationary solution.
In non-stationary flows, the separation and merger of these eddies and boundary layers become important. As \cite{sd00} notes, the lid-driven cavity flow exhibits the full spectrum of fluid mechanical phenomena within the simplest geometrical configuration.
This naturally raises the question: How can we systematically understand the flow variations and transitions through the dynamics of these prominent eddies? As we show in this paper, TFDA provides a systematic and mathematically rigorous framework to achieve this goal for dynamic flow evolutions.

We consider a two-dimensional lid-driven cavity flow in a square cavity. The movement of the lid is regularised around the corners.  We adopt the regularisation proposed in \cite{shen91}. 
Such a regularisation may not be necessary if the corner singularities are properly treated \cite{b67,  bp98, sd00}. Our adoption of regularisation is for the sake of numerical simplicity.
We specifically consider the incompressible Navier--Stokes equations, 
\begin{align}
    \partial_t {\bm u}
    + ({\bm u}\cdot \nabla) {\bm u}
    = 
    - \nabla p
    + \nu \nabla^2 {\bm u}, 
    \quad 
    \nabla \cdot {\bm u} = 0,
\end{align}
in the square domain $\Omega=\{(x,y) \in \mathbb{R}^2 \vert -1 \le x \le 1, -1 \le y \le 1\}$. 
Hence, the side length of the square is $L = 2$.
Here ${\bm u}(x, y, t)$ is the velocity field, $p$ is the pressure
and $\nu$ is the kinematic viscosity. The fluid density is normalised to one.
The boundary condition is ${\bm u} = {\bm 0}$ on the boundary walls,
except for the lid at the top. The velocity of the lid is given by
\begin{align}
    u(x, y=1, t) &= (x + 1)^2 (x - 1)^2, \\
    v(x, y=1, t) &= 0
\end{align}
where $u(x, y, t)$ and $v(x, y, t)$ are the $x$ and $y$ components of the velocity \cite{shen91}. 
Note that the velocity of the lid smoothly decreases to zero at the corners ($x=\pm 1$). 
The Reynolds number here is defined with the maximum velocity of the lid, $U= \max_{-1 \le x \le 1} u(x, y=1, t) = 1$, and the side length of the square, $L=2$, as
\begin{align}
Re = \frac{UL}{\nu}.
\label{Re}
\end{align}

The numerical scheme we use is the Chebyshev-tau method in the streamfunction and vorticity formulation of the Navier-Stokes equations. 
We use the $3/2$ rule to remove the dealiasing error. 
The number of Gauss--Lobatto collocation points is $65^2$. The boundary condition is numerically satisfied by using the influence matrix method \cite{peyret02}. 
Because of the well-known loss of the rank of the influence matrix, we need to set the vorticity to zero at four collocation points on the boundary (in addition to the corner points). 
We follow the choice of the four points proposed in \cite{ep89}, namely the nearest collocation points to the corner points on the lid and bottom wall.
The time marching scheme is semi-implicit. Specifically, we use the second-order Adams-Bashforth method for the nonlinear term and the second-order backward differentiation for the viscous term, which is known as AB/BDI2 scheme \cite{peyret02}. 
The time-step value is $\Delta t = 1.0\times 10^{-3}$.

We numerically simulate the flow for the Reynolds numbers from $Re=14000$ to $Re=16000$ in steps of $50$. 
For each $Re$, we start the simulation with the zero velocity field and continue until time $T_s \gg 1$ when the flow reaches a statistically stationary state to avoid initial transients. Then we start to analyse the flow for $T_s \le t \le T_s + \tilde{t}$. 
Their values are $T_s = 5000 = 2500 (L / U)$ and $\tilde{t}$ varying from $200 = 100 (L / U)$ to $10000 = 5000(L/U)$. 
In Table \ref{table:param}, we list numerical parameters for three representative Reynolds numbers.  The convergence study is carried out for $Re = 15000$ and $Re=16000$ by increasing the collocation points to $129^2$ or $257^2$ and halving the time step $\Delta t$ to verify that the results do not change.
 
\begin{figure}
\begin{center}
\includegraphics[width=0.45\textwidth]{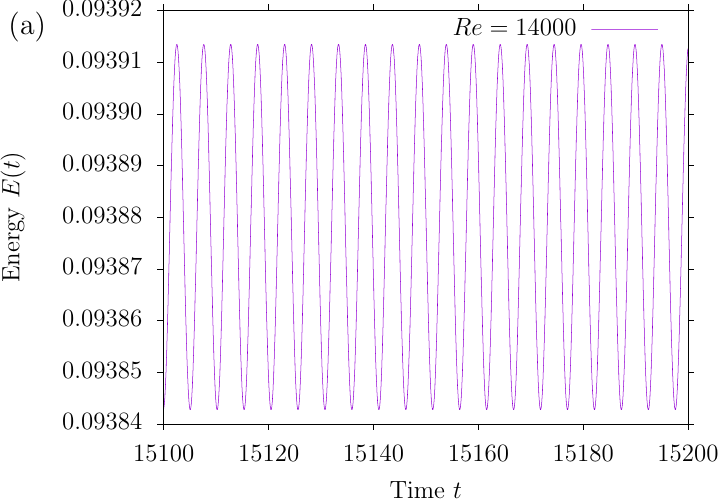}
\includegraphics[width=0.45\textwidth]{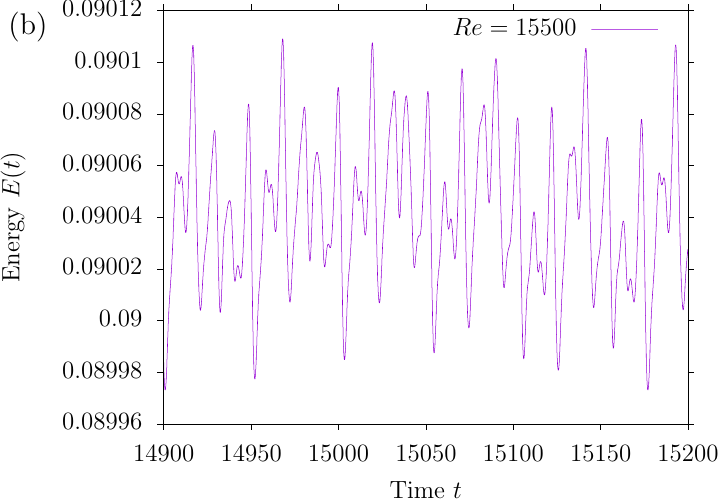}
\includegraphics[width=0.45\textwidth]{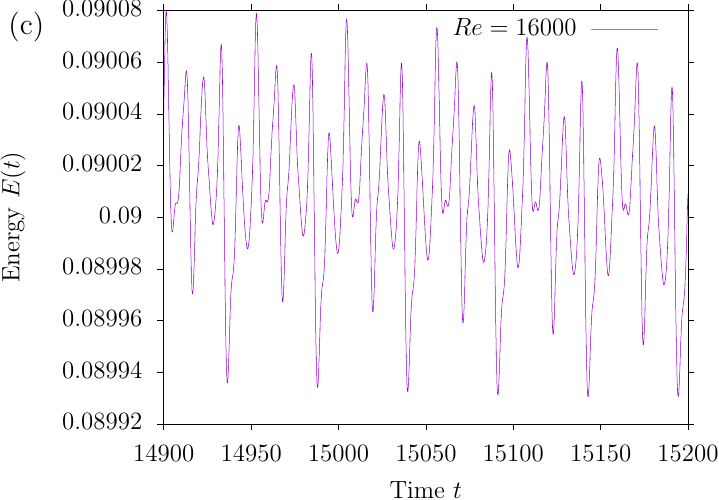}
\end{center}
\caption{\label{fig:e} The kinetic energy as a function of time for the three representative Reynolds numbers for (a) $Re=14000$, (b) $Re=15500$, and (c) $Re=16000$.}
\end{figure}

\begin{figure}
\begin{center}
\includegraphics[width=0.45\textwidth]{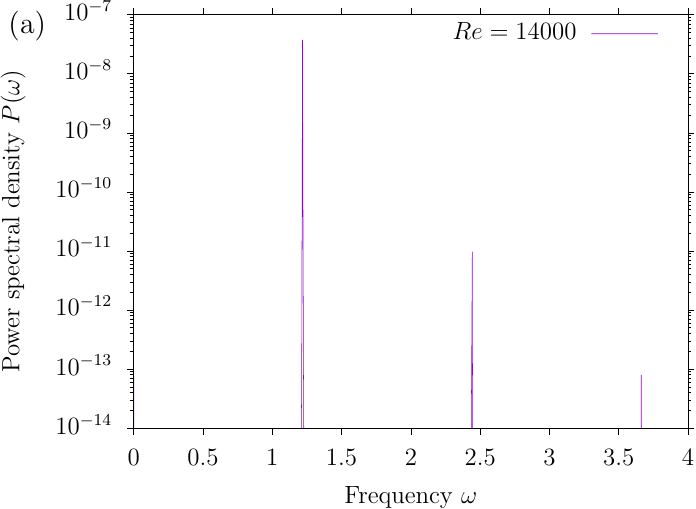}
\includegraphics[width=0.45\textwidth]{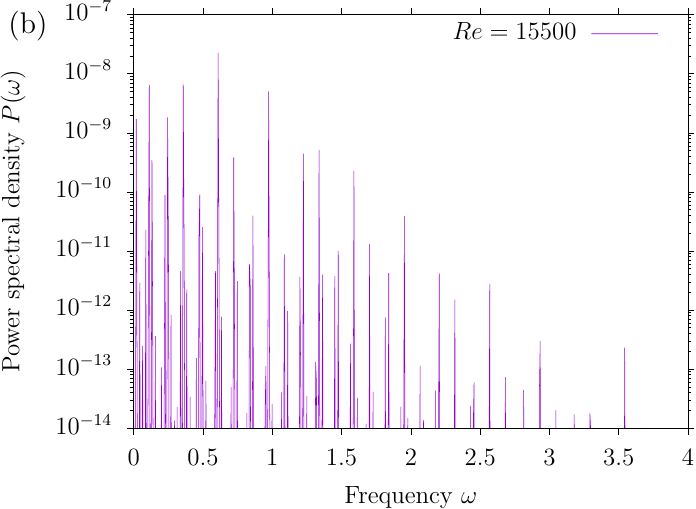}
\includegraphics[width=0.45\textwidth]{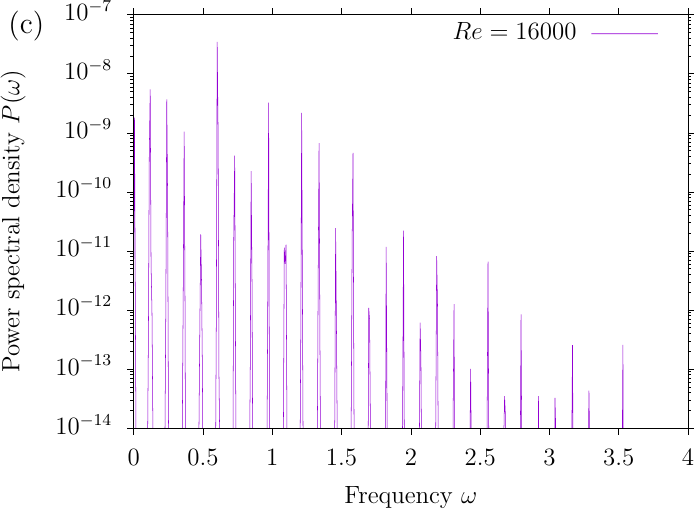}
\end{center}
\caption{\label{fig:spc} Power spectral densities of the kinetic energy for the three representative Reynolds numbers. (a) $Re=14000$, (b) $Re=15500$, and $Re=16000$. The width of the bins for the frequency is the same for the three cases. The densities are calculated from the energy $E(t)$ in $5200 \le t \le 15200$ in the step of $0.01$.}
\end{figure}

As a global fluid mechanical quantity, we consider the kinetic energy, 
\begin{align}
 E(t) = \int_{\Omega}  \frac{1}{2} |{\bm u}(x, y, t)|^2 ~{\mathrm d}x {\mathrm d}y,
 \label{defE}
\end{align} 
and use it as an indicator of the flow state as in \cite{shen91}. 
We show the variation of the kinetic energy in Figure \ref{fig:e} and its power spectral density (PSD) in Figure \ref{fig:spc}.  
The PSD for $Re=14000$ indicates that the flow is periodic. 
For $Re=15500$, \cite{shen91} concluded that the flow is quasi-periodic. 
The PSD shown in Figure \ref{fig:spc}(b) is consistent with the conclusion. 
For $Re=16000$, the PSD becomes broad and the fine structures seen at $Re=15500$ are swallowed by the broad parts. This indicates that the flow is chaotic at $Re=16000$. 
\begin{table}
  \centering
  \begin{tabular}{lcccc}
   $Re$     & collocation points & $\Delta t$             &  State \cite{shen91} & State (present) \\ \hline
   $14000$  & $65^2$             & $1 \times 10^{-3}$     &   periodic            &  periodic     \\ 
   $15500$  & $65^2$             & $1 \times 10^{-3}$     &   quasi-periodic            &  quasi-periodic \\ 
   $16000$  & $65^2$             & $1 \times 10^{-3}$     &   N/A            &  chaotic \\ 
  \end{tabular}
  \caption{Numerical parameters and flow states for representative Reynolds numbers.
  } 
  \label{table:param}
\end{table}

\subsection{Topological Flow Data Analysis}\label{sec:2.2}
We provide a brief review of TFDA used in this paper. 
For a complete description of the theory, see~\cite{TFDA-YS12,TFDA-SY18,TFDA-UYS18}.
We consider a Hamiltonian vector field of $C^r$-class ($r \ge 1$) in a two-dimensional disk $\mathcal{D}$ satisfying the slip boundary condition. 
Then the particle orbits generated by this Hamiltonian vector field coincide with the level curves of the Hamiltonian on $\mathcal{D}$. 
Since the streamfunction of the two-dimensional incompressible vector field is the Hamiltonian of the flow, the level curves are called streamlines.  
As explained in the introduction, since the vector field can depend on time, the particle orbits advected by the vector field are no longer equivalent to the level curves of the Hamiltonian in general. 
However, TFDA deals with velocity fields and their generating particle orbits for every fixed time, i.e., instantaneous streamlines.

Here, we restrict our attention to a special class of vector fields, called {\it structurally stable} Hamiltonian vector fields.
By structural stability, we mean that the topological structure of instantaneous streamlines is unchanged under any small perturbation in the $C^r$ topology.
See~\cite{Ma2005, TFDA-YS12} for a mathematically rigorous definition of structural stability. 
The set of structurally stable Hamiltonian vector fields on the disk $\mathcal{D}$ is denoted by $\mathcal{H}^r$.
This restriction gives rise to theoretically no serious problems, since $\mathcal{H}^r$ is a generic subset of the set of all $C^r$ Hamiltonian vector fields.
That is, any Hamiltonian vector field is approximated by a sequence of structurally stable Hamiltonian vector fields in $\mathcal{H}^r$.
In addition, practically, numerical simulation of flows mostly generates structurally stable streamline patterns since it hardly resolves structurally unstable orbits due to numerical approximation errors.

\begin{figure}
\begin{center}
\includegraphics[scale=0.6]{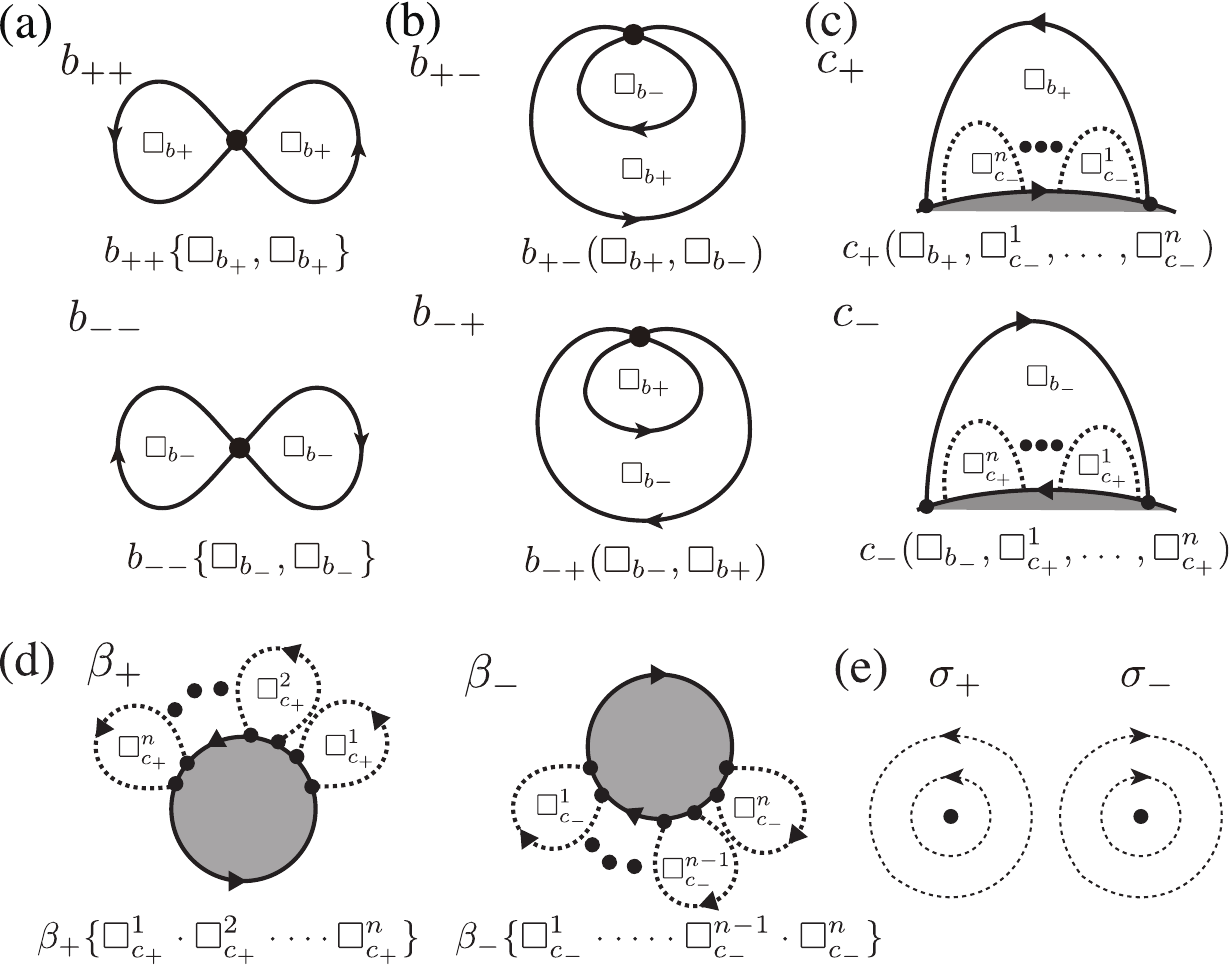}
\end{center}
 \caption{Local orbit structures appearing in the streamline patterns of structurally stable Hamiltonian vector fields in $\mathcal{D}$ and their COT symbols. 
 (a) Figure-eight patterns with a saddle and two self-connected saddle separatrices whose COT symbol is $b_{\pm\pm}\{\Box_{b_\pm},\Box_{b_\pm}\}$. 
 (b) Local orbit structures with a saddle, in which one saddle connection encloses another. The COT symbol is $b_{\pm\mp}(\Box_{b_\pm}, \Box_{b_\mp})$. 
 (c) Saddle connections between two different saddles on the same boundary. The COT symbol is $c_\pm(\Box_{b_\pm},\Box_{c_\mp}^1,\ldots,\Box_{c_\mp}^n)$. 
 (d) Isolated boundaries $\beta_\pm\{\Box_{c_\pm}^1\cdot \cdots \cdot \Box_{c_\pm}^n\}$. 
 (e) Elliptic centres $\sigma_\pm$. In each panel, $\Box_{b_\pm}$ and $\Box_{c_\pm}$ indicate that the local orbit structures in (\ref{COTsymbols}) are embedded in the flow structures as their internal structure. 
 The curly brackets $\{\}$ and the round brackets $()$ in COT symbols express whether the arrangement of the embedded structures is determined up to cyclic order and is uniquely determined respectively.}
\label{fig:local_orbits}
\end{figure}

To describe the theory of TFDA, we introduce all local orbit structures that appear in structurally stable Hamiltonian flows in $\mathcal{D}$. 
We then provide a unique letter, called a {\it COT symbol}, to each local orbit structure. Figure~\ref{fig:local_orbits}(a) shows a figure-eight orbit structure with two self-connected saddle connections. 
For these local orbit structures, we assign the COT symbols $b_{++}\{\Box_{b_+},\Box_{b_+}\}$ and $b_{--}\{\Box_{b_-},\Box_{b_-}\}$.
The subscript $\pm$ indicates the direction of the two self-connected saddle connections.
That is, we use the subscript $+$ for the saddle connection in the anticlockwise direction and the subscript $-$ for the clockwise one.
The box symbols $\Box_{b_\pm}$ represent the local orbit structures enclosed by the saddle connections, and the curly brackets $\{\}$ indicate that the arrangement of the inner local orbit structures is determined to be in cyclic order. 
The local orbit structures in $\Box_{b_\pm}$ can be chosen as follows.
\begin{equation}
 \Box_{b_\pm} \in \{ b_{\pm\pm}, b_{\pm\mp}, \sigma_\pm, \beta_\pm \}, \qquad \Box_{c_\pm} \in \{ c_\pm \}.
 \label{COTsymbols}
\end{equation}

We have other local orbital structures as in Figure~\ref{fig:local_orbits}(b) consisting of a saddle and two self-connected saddle connections, in which one saddle connection surrounds another. 
Since the flow directions along the saddle connections become the opposite, its COT symbol is given by $b_{\pm\mp}(\Box_{b_\pm}, \Box_{b_\mp})$, in which the round brackets $()$ mean that the arrangement of the inner local orbit structures is determined uniquely. 
Figure~\ref{fig:local_orbits}(c) shows local orbit structures consisting of a saddle connection between two different saddles on the same boundary.
The COT symbol $c_{\pm}(\Box_{b_\pm},\Box_{c_\mp}^1,\ldots, \Box_{c_\mp}^n)$ is assigned to the local orbit structures according to the orientations of the saddle connections, indicating that a finite number of local orbit structures $\Box_{c_\pm}^s$, $s=1,\dots, n$, chosen as (\ref{COTsymbols}), can be attached to the boundary in the anticlockwise direction if there exist. 
We also consider an isolated boundary along which the flow is going in the anticlockwise (resp. clockwise) direction in Figure~\ref{fig:local_orbits}(d).
The COT symbol is given by $\beta_+\{\Box_{c_+}^1\cdot \, \cdots \, \cdot \Box_{c_+}^n\}$ (resp. $\beta_-\{\Box_{c_-}^1\cdot \, \cdots \, \cdot \Box_{c_-}^n\}$), in which a finite number of local orbit structures are attached in the anticlockwise direction if exists, and they are arranged with the dot ``$\cdot$'' within $\{\}$ to designate the multiple cyclic order arrangement. 
 Finally, we define an isolated elliptic centre whose COT symbol is $\sigma_+$ (resp. $\sigma_-$) associated with anticlockwise (resp. clockwise) periodic orbits in Figure~\ref{fig:local_orbits}(e). Note that they have no internal structures.

\begin{figure}
\begin{center}
\includegraphics[scale=0.6]{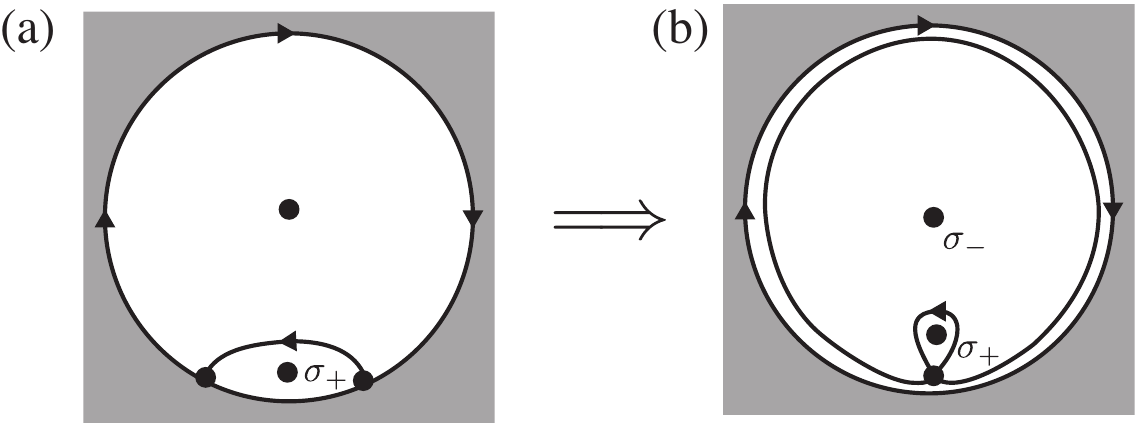}
\end{center}
  \caption{(a)An anticlockwise recirculating region on the boundary that is represented by $c_+(\sigma_+)$. (b) An isolated anticlockwise recirculating region represented by $b_{+-}(\sigma_+,\sigma_-)$.}
\label{fig:ex-separation}
\end{figure}

\begin{figure}
\begin{center}
\includegraphics[scale=0.6]{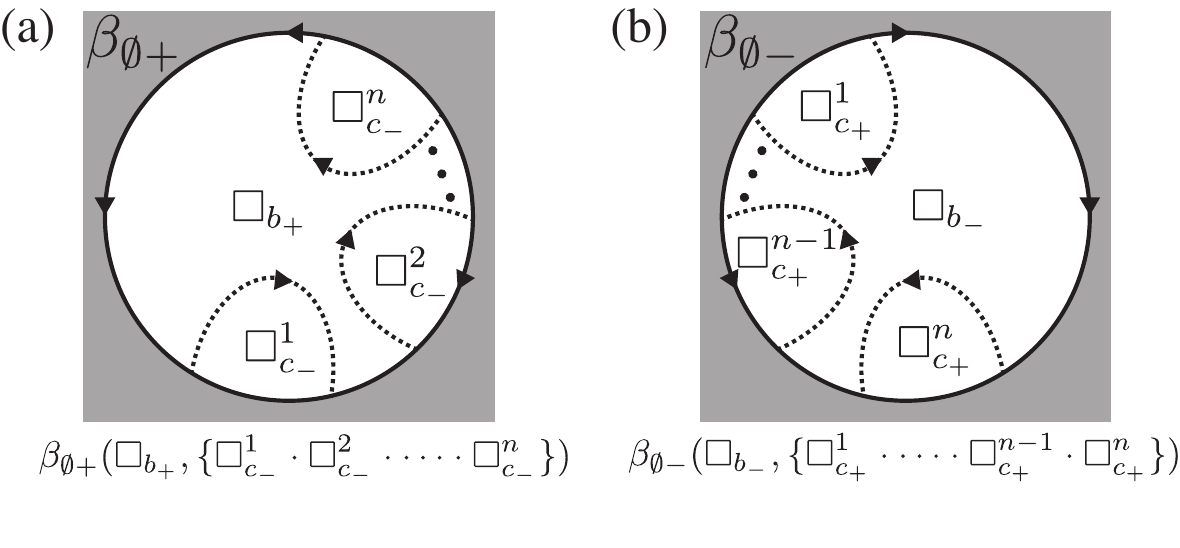}
\end{center}
\caption{Root structures in the disk $\mathcal{D}$ and their COT symbols, where a local orbit structure $b_{\pm\pm}$, $b_{\pm\mp}$, $\sigma_\pm$ or $\beta_\pm$ is embedded at $\Box_{b_\pm}$ and any number of $c_\pm$ structures can be attached along the boundary of the disk as in $\Box_{c_\pm}^s$, $s=1,\dots, n$ in the anticlockwise cyclic order if exist.
  (a) The root structure $\beta_{\emptyset+}$. 
  The flow along the boundary is going anticlockwise direction. 
  (b) The root structure $\beta_{\emptyset-}$. 
  The flow along the boundary is going clockwise direction.}
\label{fig:root}
\end{figure}
 
For the topological orbit structures generated by structurally stable Hamiltonian vector fields in $\mathcal{D}$, the following result is shown by~\cite{TFDA-YS12}.
 \begin{theorem}
 The topological structure of {\it instantaneous streamlines} generated by a structurally stable Hamiltonian vector field in $\mathcal{D}$ is uniquely represented by a combination of local orbit structures $b_{\pm\pm}$, $b_{\pm\mp}$, $c_{\pm}$,  $\sigma_{\pm}$ and $\beta_\pm$.
 \end{theorem}
We note that a change of the COT symbols can provide specific hydrodynamic information. 
For example, an anticlockwise boundary saddle represented by $c_+(\sigma_+)$ in Figure~\ref{fig:ex-separation}(a) can transition to a pattern with a homoclinic saddle expressed as $b_{+-}(\sigma_+,\sigma_-)$ in Figure~\ref{fig:ex-separation}(b). 
This transition represents a physical mechanism where an anticlockwise vortical flow region, confined to the boundary, separates from the boundary and enters the interior of the fluid region, i.e., a vortex separation. 
More examples are demonstrated in~\cite{ssy14, TFDA-SY15}.

Based on the classification theorem, we convert the topological structure of orbit connections for a given structurally stable Hamiltonian vector field in  $\mathcal{D}$  into a planar tree (COT) and its string expression (COT representation), which are the output of TFDA. 
The conversion algorithm to COT is described below. For a detailed description of the algorithm, see \Cref{sec:psiclone} and \cite{TFDA-UYS18}.
Let us first consider two topologically simplest flows in the disk $\mathcal{D}$ as shown in Figure~\ref{fig:root}, which are called root structures. 
When the flow direction along the boundary of the disk is anticlockwise (resp. clockwise), we assign the COT symbol $\beta_{\emptyset +}(\Box_{b_+}, \{\Box_{c_-}^1 \cdot  \, \cdots \, \cdot\Box_{c_-}^n\})$ (resp. $\beta_{\emptyset-}(\Box_{b_-},\{\Box_{c_+}^1\cdot  \, \cdots \, \cdot \Box_{c_+}^n\})$).
They indicate that the root structure contains at least one local orbit structure in $\Box_{b_{\pm}}$ and any number of local orbit structures $\Box_{c_\pm}^s$, $s=1,\dots, n$ can be attached along the boundary in anticlockwise cyclic order if there exist.
Next, for the given structurally stable Hamiltonian vector field (or the Hamiltonian), starting from the root structure, we identify the internal local orbit structures corresponding to $\Box_{b_\pm}$
and $\Box_{c_\pm}^s$ inductively.
Every time we identify internal local orbit structures, we create edges from the parent node of the structure containing them to the child nodes of embedded internal local structures. 
Repeating this step inductively until we reach isolated elliptic centres or boundaries with no $c_\pm$ structures attached yields a planar tree, namely a partially Cyclically Ordered rooted labelled Tree (COT).
According to~\cite{TFDA-SY18}, it has been mathematically shown that if the root node is fixed, COTs are in one-to-one correspondence with the topological orbit structures of structurally stable Hamiltonian vector fields. 
In other words, each topologically different pattern of instantaneous streamlines maps to exactly one COT, and vice versa up to topological equivalence. 
This means that a COT becomes a unique identifier of the flow pattern in terms of topology. 

\begin{figure}
\begin{center}
\includegraphics[scale=0.6]{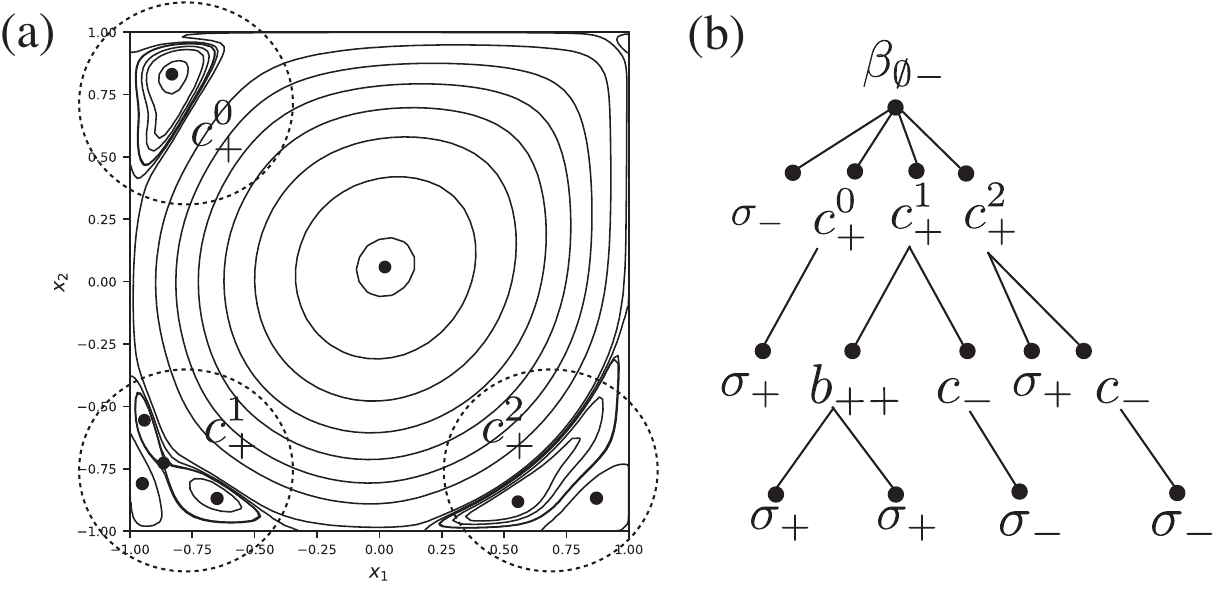}
\end{center}
  \caption{(a) A snapshot of the streamline pattern of the lid-driven cavity flow at Reynolds number $Re=14000$. (b) The partially cyclically ordered labelled rooted tree (COT) of the streamline pattern, whose COT representation is given by (\ref{ex-COT}).}
\label{fig:example-lid}
\end{figure}

\subsection{TFDA for the lid-driven cavity flow}\label{sec:2.3}

As an example, we explain how the conversion algorithm is applied to an instantaneous streamfunction in the lid-driven cavity at $Re=14000$ in Figure~\ref{fig:example-lid}(a). The flow inside the cavity can be regarded as a Hamiltonian flow in a bounded region inside a rectangle. 
 This is because the value of the streamfunction becomes a constant along the boundary of the cavity and a level curve of the same value in the open part above the lid also closes this rectangular region. Hence, the flow is topologically equivalent to a structurally stable Hamiltonian flow in the disk $\mathcal{D}$, to which the conversion algorithm is applicable. 

We show that the topological structure of the streamline pattern in Figure~\ref{fig:example-lid}(a) is converted to the COT in Figure~\ref{fig:example-lid}(b).
First, since the flow goes around the cavity boundary in the clockwise direction, the root structure of this flow is topologically identified as $\beta_{\emptyset-}$ in Figure~\ref{fig:root}(b). 
  As an internal structure of the root structure corresponding to $\Box_{b_-}$, we identify an elliptic centre $\sigma_-$ located at the centre of a large rotating flow region. 
  For orbit structures of $\Box_{c_+}^s$ along the boundary, we find three local orbit structures $c_+$ connecting between saddles at the boundary at the corners of the cavity.
  These orbit structures express the existence of the recirculating regions.
  Here, to distinguish these $c_+$ structures, 
  we tag each of them
with a corner index $i$ like $c_+^i$ that indicates which corner of the cavity it belongs to
(see Figure~\ref{fig:example-lid}(a)):
\begin{equation}
i = \begin{cases}
0 & \text{(the top left)}, \\
1 & \text{(the bottom left)}, \\
2 & \text{(the bottom right)}.
\end{cases}
\end{equation}
The top right corner index is excluded because the moving lid suppresses large persistent orbits there, as explained later. 
At any given time, a corner may host zero, one, or several $c_+$ symbols,
reflecting the intermittent birth and death of recirculation cells. 
  Since the elliptic centre $\sigma_-$ and the three $c_+$ structures are identified as the internal structures contained in the root structure $\beta_{\emptyset-}$, we create the root node of $\beta_{\emptyset -}$ and four child nodes of the local orbit structures $\sigma_-$ and $c_+^i$ for $i=0,1,2$, and then connect them with the edges. 
  Second, since the elliptic centre $\sigma_-$ in the centre of the rotating region does not have an internal structure, we stop the procedure for this node.
  Similarly, since only elliptic centres $\sigma_+$ are contained in the orbit structure $c_+^0$, the procedure is stopped after adding an edge to the node of the elliptic centre with the label $\sigma_+$ as its child node. 
  On the other hand, since an anticlockwise $b_{++}$ structure in Figure~\ref{fig:local_orbits}(a) is contained inside the $c_+^1$ structure in the bottom left corner, a new node with the label $b_{++}$ is created as a child node and an edge connecting to the parent node with the label $c_+^1$. 
  Both recirculating domains $c_+^1$ and $c_+^2$ contain a small recirculating domain $c_-$ inside, where the flow rotates in the opposite direction.
  Finally, there are isolated elliptic centres $\sigma_+$ within the orbit structure $b_{++}$ and $\sigma_-$ within the structure $c_-$, so the conversion algorithm is terminated.

The COT is compactly described as the following string, say COT representation.
  \begin{equation}
  \beta_{\emptyset-}(\sigma_-, \{c_+^0(\sigma_+)\cdot c_+^1(b_{++}\{\sigma_+, \sigma_+\},c_-(\sigma_-) )\cdot c_+^2(\sigma_+,c_-(\sigma_-))\}).
  \label{ex-COT}
  \end{equation}
The conversion from the COT of Figure~\ref{fig:example-lid}(b) to (\ref{ex-COT}) is explained below. 
We begin with the root node $\beta_{\emptyset-}$ whose COT symbol is represented by $\beta_{\emptyset-}(\Box_{b_-}, \{ \Box_{c_+}^1 \cdot \, \cdots \, \cdot \Box_{c_+}^n \})$.
Since the root node has four child nodes,
by substituting the COT symbols for the nodes into the positions of $\Box_{b_-}$ and $\Box_{c_+}^s$ for $s=1,2,3$ as $\Box_{b_-}=\sigma_{-}$, $\Box_{c_+}^1=c_+^0(\Box_{b_+},\Box_{c_-}^1 ,\ldots,\Box_{c_-}^n)$, $\Box_{c_+}^2=c_+^1(\Box_{b_+},\Box_{c_-}^1 ,\ldots,\Box_{c_-}^n)$ and $\Box_{c_+}^3 = c_+^2(\Box_{b_+},\Box_{c_-}^1 ,\ldots,\Box_{c_-}^n)$, we obtain 
$$
\beta_{\emptyset-}(\sigma_-, \{ c_+^0(\Box_{b_+},\Box_{c_-}^1,\ldots,\Box_{c_-}^n) \cdot c_+^1(\Box_{b_+},\Box_{c_-}^1 ,\ldots,\Box_{c_-}^n) \cdot c_+^2(\Box_{b_+}, \Box_{c_-}^1 ,\ldots,\Box_{c_-}^n) \}).
$$
Since the node $c_+^1$ has two nodes $b_{++}$ and $c_-$, we substitute $\Box_{b_+}=b_{++}\{\Box_{b_+}, \Box_{b_+}\}$ and $\Box_{c_-}^1 = c_-(\Box_{b_-})$ in the COT symbol $c_+^1(\Box_{b_+},\Box_{c_-}^1 ,\ldots,\Box_{c_-}^n)$ with $n=1$. 
Similarly, for the other nodes $c_+^i$ ($i=0,2)$, we substitute the COT symbols in $\Box_{b_\pm}$ and $\Box_{c_\pm}$ into the round brackets that follow them according to their respective order.
\[
 \beta_{\emptyset-}(\sigma_-, \{ c_+^0(\sigma_+)\cdot c_+^1(b_{++}\{\Box_{b_+}, \Box_{b_+}\},c_-(\Box_{b_-}) )\cdot c_+^2(\sigma_+,c_-(\Box_{b_-}))\}).
\]  
Finally, since $\sigma_\pm$ appear below each of $b_{++}$ and $c_{-}$ in the COT, substituting $\sigma_\pm$ in the corresponding to the box symbols $\Box_{b_\pm}$, we obtain (\ref{ex-COT}) as desired.

In this paper, the computation of COT and COT representations is carried out with \texttt{psiclone}, a Python library that implements the TFDA algorithm proposed in \cite{TFDA-UYS18}.
 A two-dimensional array of streamfunction values, $\psi(x,y)$, on a rectangular grid is provided as input. 
\texttt{Psiclone} extracts critical values of $\psi(x, y)$ that correspond to level sets containing saddles or centres. 
It is worth noting that this extraction does not require the direct calculation of separatrices, saddles, and centres. 
Instead, it leverages the fact that the topology of connected components enclosed by the level set changes just below and above these critical values. 
Indeed, persistent homolog, a technique from Topological Data Analysis (TDA), can discover all such critical values. Furthermore, this method incorporates the streamfunction values in topological change events, enabling the identification of noise-like structures if the difference in these critical values is small. 
In the present analysis, we keep only those topological structures that remain unchanged under perturbations of the streamfunction with amplitude $\Delta \psi < \varepsilon$ ($\varepsilon=5.0\times10^{-4}$), thus discarding topological structures caused by small changes. In spirit, this plays the role of a low-pass filter, but the mechanism respects the robustness to topological structures, rather than frequency filtering. See \Cref{sec:psiclone} for a technical review of the algorithm.

\section{Topological flow data analysis for $Re=14000$}\label{sec:3}
\subsection{Transition of streamline topology}\label{sec:3.1.1}
 Considering the lid-driven cavity flow at $Re=14000$, we describe how the topological structures of the streamline patterns change using COT representations.
  Although the flow pattern evolves continuously over time, their topological structures remain unchanged for a certain period, since they are robust under small continuous perturbations. 
  By TFDA, we show that the evolution of the flow patterns is decomposed into a discrete transition among several topological equivalence classes of flow patterns with the same COT representations.
In this analysis, we sample the streamfunction data every $10$ step after $t=T_s$, that is, at $t=t_k \equiv T_s+ 10\times k\Delta t$ for $k=0,1,\ldots, 580$.
\begin{figure}
\begin{center}
\includegraphics[scale=0.55]{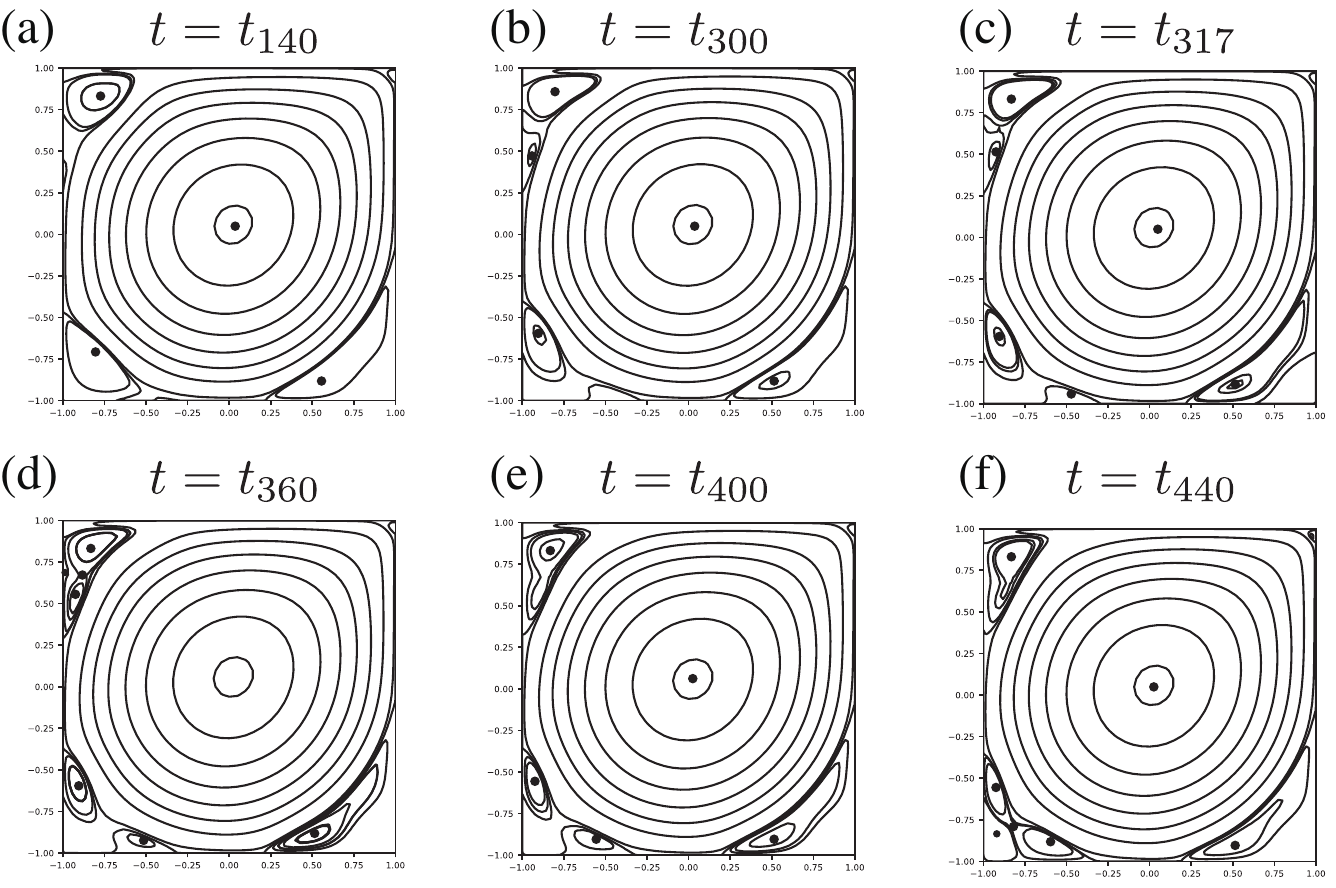}
\end{center}
\caption{Transition of the streamline patterns in the lid-driven cavity flow from $t=t_{140}$ to {$t=t_{440}$}. The Reynolds number is $Re=14000$.}
\label{fig:one_cycle_14000}
\end{figure}

Figure~\ref{fig:one_cycle_14000} shows snapshots of the evolution of the flow from $t=t_{140}$ to {$t=t_{440}$}.
{We observe six topologically different flow patterns whose COT representations are provided in Table~\ref{tbl:COT_one_cycle}.}
The COT representation is unique for flow patterns with the same topological structure, and this state continues for a certain period of evolution. Hence, each COT representation is associated with a duration, as shown in Table~\ref{tbl:COT_one_cycle}.
For instance, the flow patterns from $t=t_0$ to $t=t_{277}$ have the same topological structure as that of Figures~\ref{fig:one_cycle_14000}(a), so the COT representation $\beta_{\emptyset-}(\sigma_-, \{c_+^0(\sigma_+)\cdot c_+^1(\sigma_+)\cdot c_+^2(\sigma_+)\})$ remains unchanged in this period.
At $t=t_{278}$, the topological structure of the flow pattern transitions to the one with the same COT representation as that of Figure~\ref{fig:one_cycle_14000}(b), i.e. $\beta_{\emptyset-}(\sigma_-, \{c_+^0(\sigma_+) \cdot c_+^0(\sigma_+) \cdot c_+^1(\sigma_+) \cdot c_+^2(\sigma_+)\})$.
The same COT representation continues until {$t=t_{316}$}.
Hence, the flow patterns in Figure~\ref{fig:one_cycle_14000} are considered representative elements of each topologically equivalent class, and therefore the evolution of the flow patterns is reduced to a discrete transition between the COT representations.
We note that a tiny boundary saddle separatrix always exists in the top right corner of each panel in Figure~\ref{fig:one_cycle_14000}. 
However, this structure is disregarded by \texttt{psiclone}. 
Since the streamfunction values of the boundary separatrix and of the elliptic centre inside are close enough to be less than the threshold $\varepsilon=5.0\times10^{-4}$, the structure is considered a topologically inessential structure, as explained in the previous section. 
In addition, we observe small boundary saddle separatrices on the left wall in $t=t_{140}$ (Figure~\ref{fig:one_cycle_14000}(a)) and on the bottom wall in $t=t_{300}$ (Figure~\ref{fig:one_cycle_14000}(b)).
They are also considered small structures at these times.
However, at later times, the difference in the level values of the streamfunction for them becomes larger than the threshold $\varepsilon$ and they are eventually recognised as boundary saddle separatrices as shown in Figures~\ref{fig:one_cycle_14000}(b,c).

\begin{table}
\begin{center}
\begin{adjustbox}{scale=0.8}
\begin{tabular}{ccccc} \hline
{ID} &Figure & time & COT  & duration \\ \hline\hline
1 & \ref{fig:one_cycle_14000}(a) & $t_{140}$ & $\beta_{\emptyset-}(\sigma_-, \{ c_+^0(\sigma_+)\cdot c_+^1(\sigma_+)\cdot c_+^2(\sigma_+)\})$ & $[t_0, t_{277}]$ \\
2 & \ref{fig:one_cycle_14000}(b) & $t_{300}$ & $\beta_{\emptyset-}(\sigma_-, \{c_+^0(\sigma_+)\cdot c_+^0(\sigma_+)\cdot c_+^1(\sigma_+)\cdot  c_+^2(\sigma_+)\})$ & {$[t_{278},t_{316}]$} \\
3 & \ref{fig:one_cycle_14000}(c) & $t_{317}$ & {$\beta_{\emptyset-}(\sigma_-, \{c_+^0(\sigma_+)\cdot c_+^0(\sigma_+)\cdot c_+^1(\sigma_+) \cdot c_+^1(\sigma_+) \cdot c_+^2(\sigma_+)\})$} & 
{$t_{317}$}\\
4 &\ref{fig:one_cycle_14000}(d) & $t_{360}$ & $\beta_{\emptyset-}(\sigma_-, \{c_+^0(b_{++}\{\sigma_+, \sigma_+\}, c_-(\sigma_-))\cdot c_+^1(\sigma_+)\cdot c_+^1(\sigma_+)\cdot c_+^2(\sigma_+)\})$ & $[t_{318},t_{397}]$\\
5 &\ref{fig:one_cycle_14000}(e) & $t_{400}$ & $\beta_{\emptyset-}(\sigma_-, \{c_+^0(\sigma_+)\cdot c_+^1(\sigma_+)\cdot c_+^1(\sigma_+)\cdot c_+^2(\sigma_+)\})$ & $[t_{398},t_{424}]$\\
6 &\ref{fig:one_cycle_14000}(f) & $t_{440}$ & $\beta_{\emptyset-}(\sigma_-, \{c_+^0(\sigma_+)\cdot c_+^1(b_{++}\{\sigma_+, \sigma_+\}, c_-(\sigma_-)) \cdot c_+^2(\sigma_+)\})$ & $[t_{425},t_{497}]$\\
- & - &  - & $\beta_{\emptyset-}(\sigma_-,\{ c_+^0(\sigma_+)\cdot c_+^1(\sigma_+)\cdot c_+^2(\sigma_+)\})$ & $[t_{498},t_{580}]$ \\ \hline
\end{tabular}
\end{adjustbox}
\end{center}
\caption{The COT representations for the flow patterns in Figure~\ref{fig:one_cycle_14000}. }
\label{tbl:COT_one_cycle}
\end{table}

Based on the COT representations in Table~\ref{tbl:COT_one_cycle}, we describe how the topological structure of the flow changes during this period. 
 As mentioned in the previous section, \texttt{psiclone} can compute the COT representation for a given streamfunction without numerically determining centres, saddles, or separatrices. 
 On the other hand, in Figure~\ref{fig:one_cycle_14000}, we plot critical points using markers when the difference in the level values of the streamfunction for them becomes greater than the threshold $\varepsilon$, and we plot separatrices for reference. 
 All COT representations start from the same COT symbol $\beta_{\emptyset-}$ whose first component is $\sigma_-$ indicating the existence of an isolated elliptic fixed point at the centre of the large clockwise rotational flow in the cavity.
 Hence, we pay attention to the local flow structure represented by the sequence of COT symbols inside the curly brackets of the COT symbol for $\beta_{\emptyset-}$. 
 In the COT representation for Figure~\ref{fig:one_cycle_14000}(a) at $t=t_{140}$, there are the three COT symbols $c_+^i(\sigma_+)$, $i=0,1,2$ inside the curly brackets.
 Each symbol represents an anticlockwise recirculating flow region enclosed by a boundary saddle connection in the corner of the cavity.
  Here, we focus on the change in the COT symbols for $c_+^0$. 
  At $t=t_{300}$, another $c_+^0(\sigma_+)$ is added to the COT representation for the pattern of Figure~\ref{fig:one_cycle_14000}(b),  indicating that a new boundary saddle connection appears on the left wall in the region of the top left corner. 
{ With a brief fluctuation in $t=t_{317}$ with respect to the bottom left corner as shown in Figure~\ref{fig:one_cycle_14000}(c), the pair of COT symbols $c_+^0(\sigma_+)\cdot c_+^0(\sigma_+)$ in $t=t_{300}$ is then replaced by $c_+^0(b_{++}\{\sigma_+, \sigma_+\}, c_-(\sigma_-))$ in $t=t_{360}$.}
  This means that the two boundary saddle connections merge and transition to a local orbit structure of a boundary saddle connection that contains a figure-eight orbit structure as observed in the top left corner of Figure~\ref{fig:one_cycle_14000}(d). 
  Subsequently, $c_+^0(b_{++}\{\sigma_+,\sigma_+\}, c_-(\sigma_-))$ is reduced to a simple $c_+^0(\sigma_+)$ at $t=t_{400}$.
  Hence, the figure-eight orbit structure disappears and transforms into a simple boundary saddle connection as in Figure~\ref{fig:one_cycle_14000}(d). 
  The COT symbol $c_+^1(\sigma_+)$ for the bottom left corner eddy at $t=t_{300}$ changes similarly as $c_+^1(\sigma_+)\cdot c_+^1(\sigma_+) \rightarrow c_+^1(b_{++}\{\sigma_+, \sigma_+\}, c_-(\sigma_-)) \rightarrow c_+^1(\sigma_+)$ from {$t=t_{317}$} to $t=t_{580}$. 
  This also indicates that the local orbit structure in the bottom left corner is subject to the same topological transitions as that in the top left corner. 
  {After these transitions occur, the COT representation finally returns the same COT representation at $t=t_{498}$ as that of $t=t_{140}$, and it continues until $t=t_{580}$.}
  
\begin{figure}
\begin{center}
\includegraphics[width=12cm]{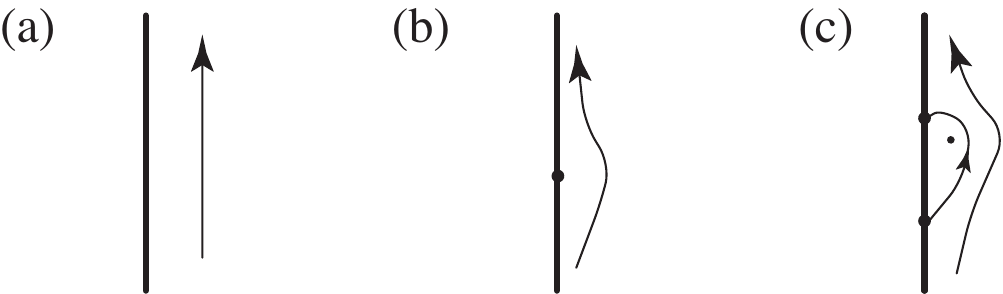}
\end{center}
\caption{Generation of a local orbit structure represented by $c_+(\sigma_+)$ on the left wall through a pinching transition.}
\label{fig:p-unstable}
\end{figure}

\begin{figure}
\begin{center}
\includegraphics[width=12cm]{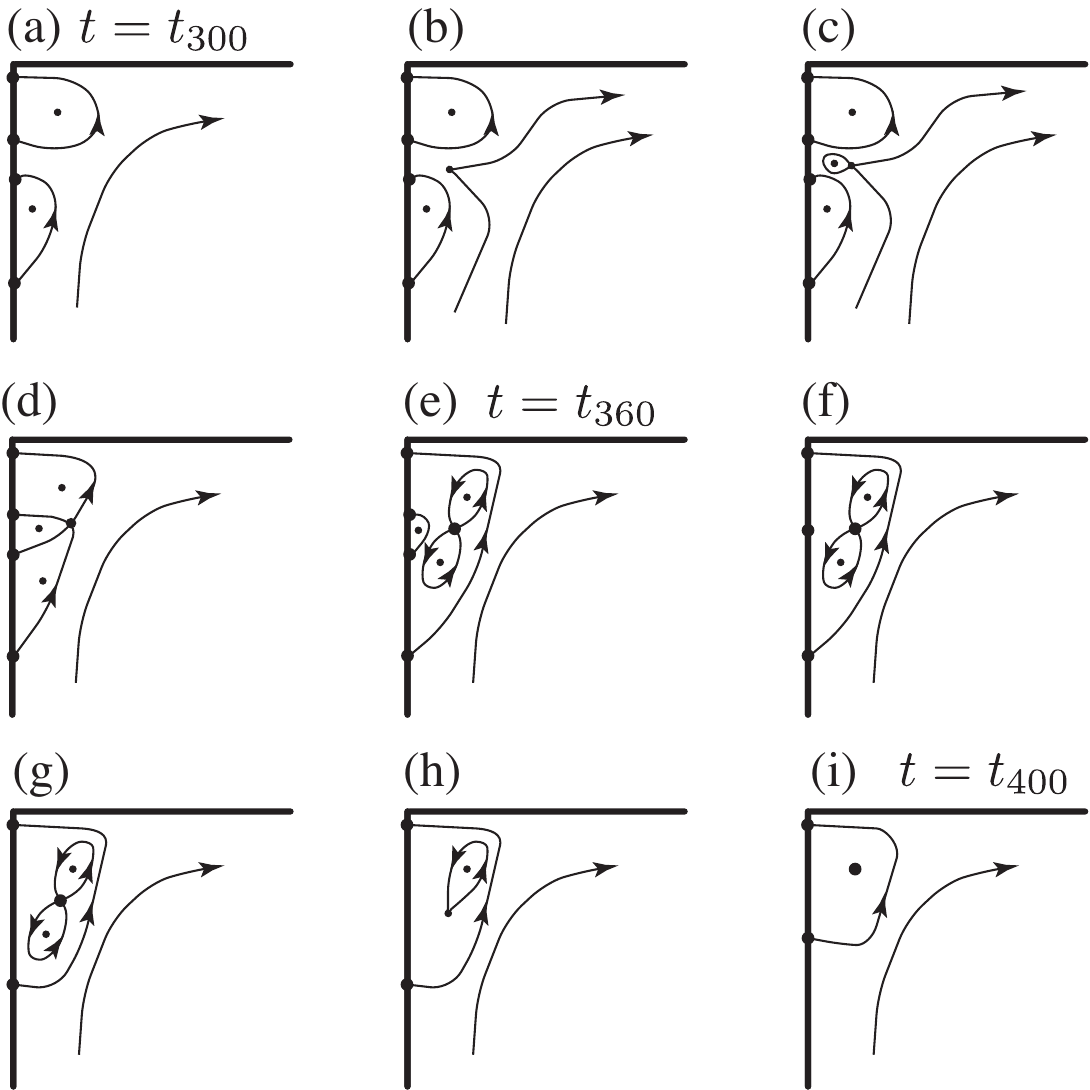}
\end{center}
  \caption{Schematic of topological transitions for $Re=14000$ of local flow patterns in the top left corner from $t=t_{300}$ to $t=t_{400}$ that is interpolated based on the topological transition theory~\cite{TFDA-SY15}.
  (a) The flow pattern at $t=t_{300}$ represented by $c_+^0(\sigma_+)\cdot c_+^0(\sigma_+)$.
  (b) A singular orbit with a degenerate pinching emerges. 
  (c) Creation of a local self-connected saddle connection between the two boundary saddle connections through the pinching. 
  This is structurally stable, but it is not detected by TFDA from the present dataset. 
  (d) A structurally unstable flow pattern with heteroclinic connections. (e) The structurally stable flow pattern observed at $t=t_{360}$ represented by $c_+(b_{++}\{\sigma_+, \sigma_+\}, c_-(\sigma_-))$. (f) The inner clockwise boundary saddle connection $c_-(\sigma_-)$ disappears through the pinching transition on the left wall as shown in Figure~\ref{fig:p-unstable}. (g) A structurally stable represented by $c_+(b_{++}\{\sigma_+, \sigma_+\})$ that is not detected in the dataset by TFDA, either.
  (h) The local orbit structure $b_{++}$ disappears through a pinching transition. (i) The structurally stable pattern  represented by $c_+(\sigma_+)$ at $t=t_{400}$.}
\label{fig:h-unstable}
\end{figure}

\cite{TFDA-SY15} proposed a theory to interpolate topological pattern transitions that can occur from changes in COT symbols. 
Based on this theory, it is possible to describe an achievable pathway in the region of the top left corner even if the time series data do not contain intermediate patterns of the transition.
First, the generation of a boundary saddle connection $c_+^0(\sigma_+)$ on the left wall in the region of the top left corner from $t=t_{140}$ to $t=t_{300}$ is schematically provided in Figure~\ref{fig:p-unstable}. 
From Figure~\ref{fig:p-unstable}(a) to Figure~\ref{fig:p-unstable}(c), a degenerate boundary stagnation point appears on the left wall as in Figure~\ref{fig:p-unstable}(b) and divides into two regular boundary saddles.
This transition is called a {\it pinching transition}. 
The flow of Figure~\ref{fig:p-unstable}(b) is hardly observed in the numerical simulation, since it is structurally unstable. 
Next, the transition of the boundary saddle connection in the top left corner from $t=t_{300}$ to $t=t_{400}$ is expressed as the following changes in the COT symbols.
  \begin{equation}\label{transition-h-untable}
  c_+^0(\sigma_+)\cdot c_+^0(\sigma_+) \longrightarrow c_+^0(b_{++}\{\sigma_+, \sigma_+\}, c_-(\sigma_-)) \longrightarrow c_+^0(\sigma_+).
  \end{equation}
 An interpolated transition pathway is shown schematically in Figure~\ref{fig:h-unstable}.
 From $t=t_{300}$ to $t=t_{360}$, the two boundary saddle connections represented by $c_+^0(\sigma_+)\cdot c_+^0(\sigma_+)$ in Figure~\ref{fig:h-unstable}(a) approach, and a self-connected saddle connection between them (Figure~\ref{fig:h-unstable}(c)) emerges through a pinching transition (Figure~\ref{fig:h-unstable}(b)). 
 These local orbit structures merge to form a structurally unstable pattern with heteroclinic connections (Figure~\ref{fig:h-unstable}(d)).
 Note that in the present TFDA, no COT representation corresponding to the pattern of Figure~\ref{fig:h-unstable}(c) is observed. 
 This indicates that this transition occurs very quickly, and so streamfunction with this pattern was not contained in the time-series data. 
 However, even in this situation, the transition theory can propose this route as one of the possible transitions.  
 A slight perturbation of Figure~\ref{fig:h-unstable}(d) instantaneously gives rise to the transition to the same structurally stable pattern with a figure-eight orbit structure as the pattern at $t=t_{360}$ in Figure~\ref{fig:h-unstable}(e). 
 This transition is called {\it a heteroclinic transition}. 
 The local orbit structure after the heteroclinic transition is represented by COT symbols $c_+^0(b_{++}\{\sigma_+, \sigma_+\}, c_-(\sigma_-))$, indicating that a small clockwise recirculating flow region $c_-(\sigma_-)$ on the left wall is enclosed by the anticlockwise saddle connection $c_+^0$. 
From $t=t_{360}$ to $t=t_{400}$, we need two more steps to complete the transition.
 The small clockwise recirculating flow $c_-(\sigma_-)$ disappears due to the reverse pinching transition on the left wall (Figure~\ref{fig:h-unstable}(f)).
 Consequently, this produces a structurally stable flow pattern as in Figure~\ref{fig:h-unstable}(g).
 This flow pattern is also not detected by TFDA because it persists for a short time in the evolution.
 Finally, the elliptic centre and the saddle in the figure-eight pattern merge by a pinching transition (Figure~\ref{fig:h-unstable}(h)), and the flow pattern transfers to a simple local orbit structure of $c_+^0(\sigma_+)$ like Figure~\ref{fig:h-unstable}(i) at $t=t_{400}$.
 
{
It is important to note that a substantial body of literature has been devoted to the study of topological bifurcations of this kind from analytical perspectives; see, e.g., \cite{brons1999streamline,brons2007streamline_ad,bakker2012bifurcations,brons2007streamline}.
These studies consider certain perturbation parameters that develop rigorous local bifurcations (e.g., bubble (or zone) creations and global bifurcations) within parameter spaces.
On the other hand, the pattern change presented here is one possible route through the minimum number of transitions suggested by topological transition theory.
Hence, to determine what transitions actually occur, it is necessary to perform TFDA on time series data sampled with a finer temporal frequency.
Our topological approach, along with that of \cite{fomenko2004integrable} in the field of integrable systems, focuses on global topological structures, with the aim of providing a comprehensive classification of all possible transitions that can occur combinatorially.}

 \subsection{Transition diagram of the flow evolution}\label{sec:3.1.2}
As we see in Section~\ref{sec:3.1.1}, the change of flow patterns from $t=t_0$ to $t=t_{580}$ is expressed as a COT representation cycle. 
{
Topologically equivalent flow patterns are represented by a single COT representation, which we refer to as a \textit{topological state}. Under this representation, the continuous evolution of the flow can be reduced to a discrete dynamical process on the set of topological states.
To construct this description, we compute the COT representation of the streamfunction at each time step during the evolution. This produces a time series of COT representations that captures the sequence of topological configurations realised by the flow. Each distinct COT representation is identified as a node corresponding to a topological state. Whenever the representation changes from one time step to the next, a directed edge is introduced from the preceding node to the subsequent one.
Through this procedure, we obtain a directed graph whose nodes represent topological states and whose edges represent transitions between them. We refer to this graph as a \textit{transition diagram}. It provides a compact visual representation of the evolution of the lid-driven cavity flow in terms of discrete topological changes.}

\begin{figure}
\begin{center}
\includegraphics[width=8cm]{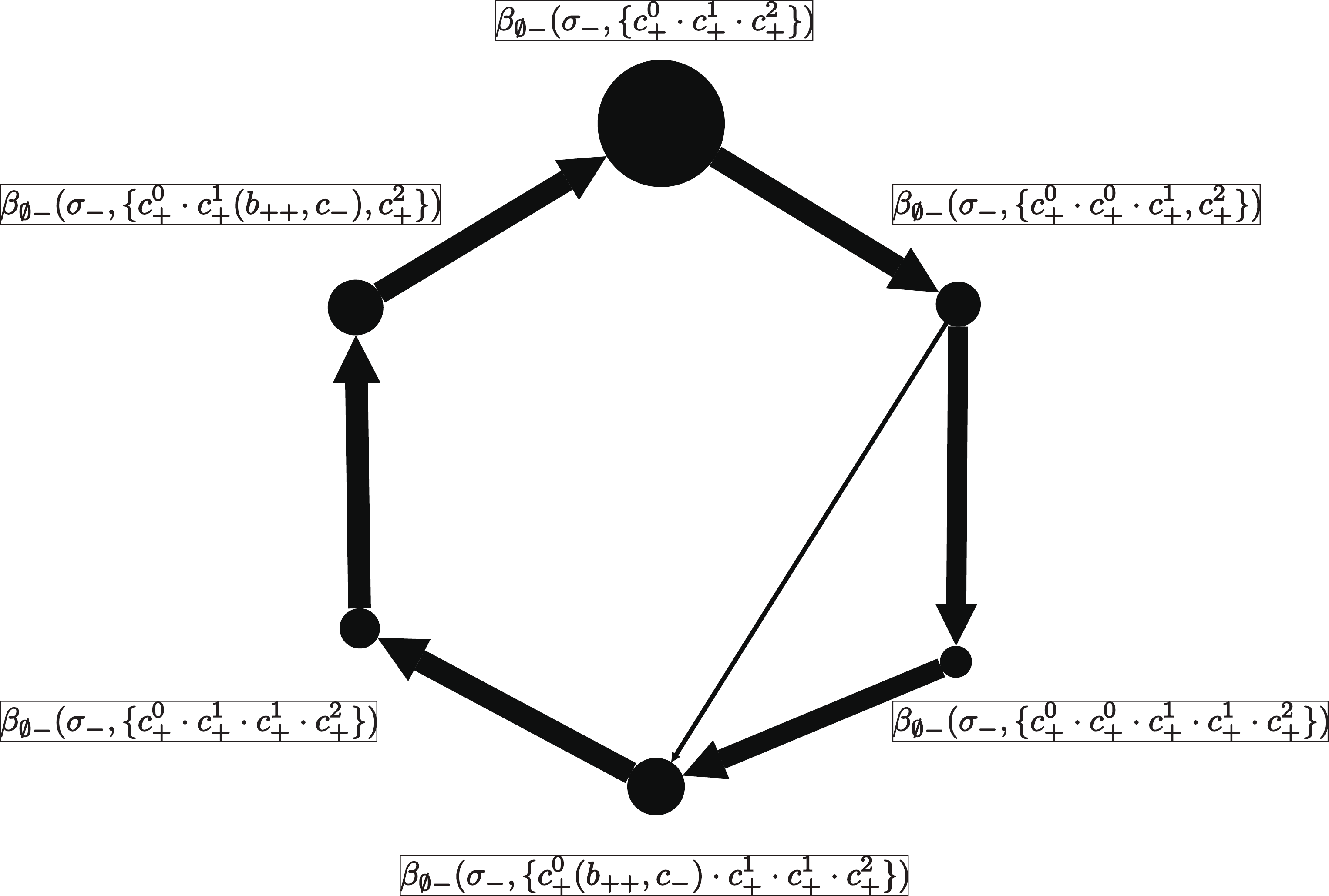}
\end{center}
  \caption{Transition diagram for the evolution of the lid-driven cavity flow for $Re=14000$.}
\label{fig:Diagram_Re14000}
\end{figure}

{Figure~\ref{fig:Diagram_Re14000} presents the transition diagram for $Re = 14000$, constructed by tracking the flow evolution over an extended time interval from $t = t_0$ to $t_{10000}$.
Each node is labelled by a simplified COT expression. For example, the label $\beta_{\emptyset-}(\sigma_-, \{c_+^0(b_{++}, c_-)\cdot c_+^1 \cdot c_+^1 \cdot c_+^2\})$ corresponds to the full COT representation $\beta_{\emptyset-}(\sigma_-, \{c_+^0(b_{++}\{\sigma_+, \sigma_+\}, c_-(\sigma_-)) \cdot c_+^1(\sigma_+) \cdot c_+^1(\sigma_+) \cdot c_+^2(\sigma_+)\})$.
We note that the topological states corresponding to these nodes are topologically equivalent to the patterns in panels (a)--(f) of \Cref{fig:one_cycle_14000}.
The size of each node in Figure \ref{fig:Diagram_Re14000} indicates the time spent on the corresponding state, while the thickness of each edge represents the number of observed transitions between the connected states. 
The diagram shows that the topological state $\beta_{\emptyset-}(\sigma_-, \{c_+^0 \cdot c_+^1 \cdot c_+^2\})$, which is the simplest topological structure as in Figure~\ref{fig:one_cycle_14000}(a), has the longest existence time in the time series.
As observed in Table~\ref{tbl:COT_one_cycle}, there is a dominant hexagonal one-way transition cycle with thick directed edges, corresponding to the transition of topological states in panels (a) $\to$ (b) $\to$ (c) $\to$ (d) $\to$ (e) $\to$ (f) $\to$ (a) of Figure~\ref{fig:one_cycle_14000}.
This constitutes the principal transition route.
In addition, a short-cut transition is visible, connecting $\beta_{\emptyset-}(\sigma_-, \{c_+^0 \cdot c_+^0 \cdot c_+^1 \cdot c_+^2\})$ directly to $\beta_{\emptyset-}(\sigma_-, \{c_+^0(b_{++}, c_-) \cdot c_+^1 \cdot c_+^1 \cdot c_+^2\})$.
This arises because the intermediate state $\beta_{\emptyset-}(\sigma_-, \{c_+^0 \cdot c_+^0 \cdot c_+^1 \cdot c_+^1 \cdot c_+^2\})$ appears for at most one frame at the present sampling rate and is occasionally missed, as indicated in Table~\ref{tbl:COT_one_cycle}.
It should also be noted that  two instances of $c_+^1$ are visually identifiable in Figure~\ref{fig:one_cycle_14000}(b), yet they fall below the adopted detection threshold, whereas in Figure~\ref{fig:one_cycle_14000}(c) they are counted as two distinct $c_+^1$ components. Therefore, we emphasize that the structure of the transition diagram depends on the threshold and the sampling frequency.}

\section{Topological flow data analysis across Reynolds numbers}\label{sec:4}
We demonstrate how TFDA extracts physical information from the lid-driven cavity flow by exploiting the time series of COT representations.
Our analysis focuses on Reynolds numbers in the range of $14000 \le Re \le 16000$, where the flow is considered to have reached a statistically stationary state at time $t = T_s = 5000$.
From this time onward, we analyse the subsequent evolution of the flow topology.
Specifically, we sample the time series of the streamfunction over the interval $0 < t - T_s \le 200$.
Snapshots are recorded every $10 \times \Delta t = 0.01$, where $\Delta t = 10^{-3}$ is the time-step size used in the direct numerical simulation.
This yields a total of $20000$ streamfunction snapshots at
$t_k = T_s + 0.01\times k$ for $k = 1, \ldots, 20000$ for each time series.
The Reynolds numbers are sampled non-uniformly within the interval:
every $250$ for $14000 \le Re \le 15250$, and every $50$ for $15300 \le Re \le 16000$,
where the finer resolution in the higher range is used to capture more detailed dynamical changes.

\input{physical_insight}

\subsection{Intrinsic causality analysis between flow patterns in corner regions}\label{sec:CCM}
\input{CCM}

\section{Summary}\label{sec:5}
TFDA is a new method of topological data analysis for two-dimensional vector fields. 
It allows us to extract all local structures to be identified geometrically and to describe their global structure as a tree (COT) and its string representation (COT representation).
Each node label (COT symbol) of the tree expresses a local orbit structure in the flow. 
The global configuration among local orbit structures is expressed as edges of the tree without ambiguity. 
Since the snapshots of the flow patterns are converted into a series of COT representations, the continuous evolution of two-dimensional incompressible flows is reduced to a discrete dynamical system between topologically equivalent flow patterns. 

We apply TFDA to the evolution of the lid-driven cavity flow for the Reynolds number ranging from $Re=14000$ to $16000$, which has been well investigated as a test bed for many existing analysis methods. 
When the Reynolds number is $Re=14000$, six topologically different streamline patterns, called topological states, are observed, and their corresponding COT representations repeat periodically throughout the evolution. 
By examining the changes in the COT representations of the flow patterns, we can theoretically identify the marginal flow pattern between them, even if it is not explicitly contained in the simulation data.
In addition, by tracking the time series of the COT representations, we can express the evolution of the flow as a graph that describes a cyclic transition between topological states.

{Using TFDA for the time series of COT representations for various Reynolds numbers, we demonstrate how one can understand the physics of the lid-driven cavity flow.
We have conducted the following five analyses.
First, we argue that the transition in flow complexity can be studied by the number of topological states identified with TFDA.
We observe critical behaviour when the Reynolds number exceeds $Re=15400$.
 Quantitative aspects, such as the value of the critical exponent, are affected by how one samples the Reynolds number, the time series of the streamfunction, and the threshold of TFDA to calculate COT. 
Second, we show that the period of periodic flow dynamics can be estimated by the time series of COT representations.
Third, we argue that the variation of physical quantities of interest, such as energy and enstrophy, can be interpreted in terms of local flow patterns of topological states and the sequence thereof. 
This can be used in addition to classifying or predicting a flow state. 
In particular, here, we focus on local maxima and minima of enstrophy and the corresponding topological states.
Fourth, we perform TFDA for flows of various Reynolds numbers, and assign serial numbers (IDs) to the topological states in the order of their appearance.
We then investigate their distribution and the rate of occurrence.
Consequently, we found that topological states with simple structures that appear in periodic motion at low Reynolds numbers remain observed for a long time in chaotic motion at high Reynolds numbers.}
Finally, we employ convergence cross mapping (CCM) to investigate causal relationships between topological flow pattern changes across different cavity corners. 
This method analyses vectorised information from the COT representations at each corner to determine how changes in one corner's recirculating domain influence another. 
In the periodic flow regime ($Re = 14000 -15300$), we do not detect significant causal relationships between pattern changes at different corners.
In the chaotic regime ($Re=15500 - 16000$), we observe asymmetric causality: changes in the topological pattern at the top left corner exhibit a stronger causal influence on the bottom left corner than vice versa.

The primary strengths of TFDA derive from two key characteristics: robustness and interpretability. Unlike traditional methods such as POD, which are sensitive to measurement noise and numerical artifacts, COT maintain their structural integrity under perturbations owing to their topological nature. This characteristic makes the analysis particularly reliable for experimental data where noise is inevitable.
A notable example of the robustness of TFDA is its application to left ventricular flows~\cite{TFDA-SI23}, where TFDA successfully identified vortex structures even in noisy clinical flow data obtained by echocardiography. 
The interpretability of COT comes from their direct connection to physically meaningful flow structures. Each symbol in a COT corresponds to a specific qualitative feature of the flow, enabling immediate physical interpretation. This attribute contrasts with POD, where the physical meaning of individual modes occasionally remains obscure.

However, on the other hand, the COT methodology faces important limitations.
First, since the streamfunction is not Galilean invariant, it should be noticed that the results obtained here with TFDA are coordinate dependent. 
Second, the current formulation is inherently based on two-dimensional topology, necessitating the projection of three-dimensional flows onto appropriate two-dimensional sections. 
Although this approach has proven effective in specific applications, including our analysis of left ventricular flow patterns \cite{TFDA-SI23}, it requires careful selection of projection planes to capture the most relevant flow feature.
This dimensional restriction, though limiting, often aligns well with practical applications where key flow features develop primarily in specific planes or where experimental measurements are naturally restricted to particular cross sections. 
Future research directions may include extending the COT framework to a fully three-dimensional topological analysis.

\appendix

\section{Linear response function analysis}\label{sec:Appendix}
To highlight the differences between observational and interventional approaches in detecting causality, we present a numerical analysis of the linear response function under spatially localised forcing. 
This analysis provides complementary insights to the convergent cross mapping results discussed in Section~\ref{sec:CCM}.
Let us recall the formalism of the linear response function \cite{PhysRepMarconi}. 
In the numerical simulation of the lid-driven cavity, the flow field reaches a statistically steady state such as a time-periodic state or a chaotic state at $t = T_s$.
Then we start adding an external forcing $f_\omega(\vec{x}, t)$ to the vorticity equation at this time $T_s$ and stop it at $t = T_s + T_f$. 
The forcing is localized in space and the duration $T_f$ is suitable, as we will discuss later.
We denote the velocity and the vorticity with the forcing by $\widetilde{\vec{u}}$ and $\widetilde{\omega}$. 
The vorticity $\widetilde{\omega}$ obeys the forced equation,
\begin{align}
\partial_t  \widetilde{\omega} + (\widetilde{\vec{u}}\cdot\nabla)\widetilde{\omega}
  = \nu \nabla^2 \widetilde{\omega} + f_\omega, \quad T_s \le t \le T_s + T_f.
 \label{forced_vor_eq}
\end{align}
We numerically solve both the forced equation (\ref{forced_vor_eq}) and the unforced vorticity equation.
The velocity and the vorticity without forcing are denoted by $\vec{u}$ and $\omega$. 
If the forcing is small enough, the difference between the two vorticity fields can be expressed with the linear response function $G_\omega(\vec{x},t \vert \vec{x}', t')$ as
\begin{align}
 \Delta \omega(\vec{x}, t) = \widetilde{\omega}(\vec{x}, t) - \omega(\vec{x}, t)
  = \int_{\Omega} {\rm d}\vec{x}' \int_{T_s}^t {\rm d} t'~ G_\omega(\vec{x}, t|\vec{x}', t') f_\omega(\vec{x}', t'),
 \quad t \ge T_s.
 \label{delta} 
\end{align}

Now, for a moment, suppose formally that the forcing is localized at $\vec{x}_f$ and $T_s$, that is, $f_\omega(\vec{x}, t) = a \delta(\vec{x} - \vec{x}_f) \delta(t - T_s)$, where  $\delta(x)$ is the Dirac delta function and $a$ is the forcing amplitude. 
Then we have
\begin{align}
a^{-1} \Delta \omega(\vec{x}, t;\vec{x}_f, T_s) =  G_\omega(\vec{x}, t| \vec{x}_f, T_s),  \quad t \ge T_s.
 \label{delta_omega}
\end{align}
With this formalism, we then consider the implication of the causation in terms of the linear response function. Let us define that $D_0$ is the region where the structure $c^0_+$ occurs (the top left corner of the cavity) and $D_1$ is the region where the structure $c^1_+$ occurs (the bottom left corner of the cavity).
The implication of $c^1_+$ being more causal than $c^0_+$ is that the forcing added in $D_1$ induces larger differences than the forcing added in $D_0$ does. 
Thus our anticipation is 
\begin{align}
 \left|\left.G_\omega(\vec{x}, t|\vec{x}_f, T_s)\right|_{\vec{x} \in D_1, \vec{x}_f \in D_0}\right|
<
 \left|\left.G_\omega(\vec{x}, t|\vec{x}_f, T_s)\right|_{\vec{x} \in D_0, \vec{x}_f \in D_1}\right|,
 \label{Grel}
\end{align}
or equivalently,
\begin{align}
 \left| \left. a^{-1}\Delta \omega(\vec{x}, t;\vec{x}_f, T_s)\right|_{\vec{x} \in D_1, \vec{x}_f \in D_0}\right|
<
 \left| \left. a^{-1}\Delta \omega(\vec{x}, t;\vec{x}_f, T_s)\right|_{\vec{x} \in D_0, \vec{x}_f \in D_1}\right|.
 \label{delta_rel}
\end{align}
The linear response functions of the streamfunction can be defined likewise, and we expect that similar inequalities hold. 
The region $D_0$ (the top left corner) is specifically defined as
\begin{align}
 D_0:~ -1 \le x \le -1/2,~  1/2 \le y \le 1,
\end{align}
and the region $D_1$ (the bottom left corner) as
\begin{align}
 D_1:~ -1 \le x \le -1/2,~  -1 \le y \le -1/2.
\end{align}
\begin{figure}
\centerline{
 \includegraphics[width=0.45\textwidth]{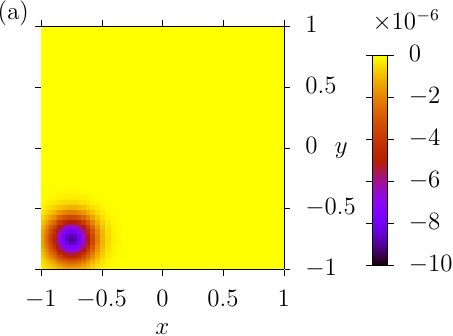}
 \hspace*{0.6cm} 
 \includegraphics[width=0.45\textwidth]{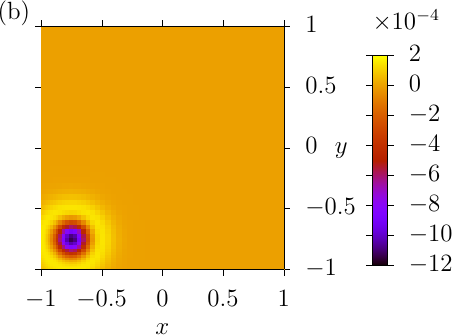} 
 } 
\caption{\label{forcing} (a) The localized forcing for the streamfunction $f_\psi(\vec{x}, t)$ (equation (\ref{fpsi})), added in the region $D_1$ ($\vec{x}_f \in D_1$).
 (b) The corresponding forcing for the vorticity $f_\omega(\vec{x}, t) = - \nabla^2 f_\psi(\vec{x}, t)$. In the bright yellow region in the panel (b), the forcing $f_\omega$ is positive. In other words, the profile of $f_\omega$ is like a Mexican hat.}
\end{figure}

In practice, we do not use the delta function for the forcing, but the Gaussian
function. 
Nonetheless, we have another problem with the locality of the forcing. 
If we use the Gaussian function for the vorticity forcing $f_\omega(\vec{x}, t)$, the corresponding forcing for the streamfunction $f_\psi(\vec{x}, t)$ becomes much less localized. 
To ensure localization for both, we use the Gaussian function for the streamfunction forcing. 
Namely,
\begin{align}
 f_\psi(\vec{x}, t)
 =
 \begin{cases}
  \displaystyle
  \frac{a}{2\pi\sigma^2}\exp\left[-\frac{1}{2}\frac{|\vec{x} - \vec{x}_f|^2}{\sigma^2}\right],  &  T_s \le t \le T_s + T_f, \\
  \displaystyle 0,  & \text{otherwise}.
  \label{fpsi}
 \end{cases}
\end{align}
Here, the width of the forcing is set to $\sigma = 1/8$. We set $\vec{x}_f$ either to the centre of $D_0$ ($(x_f, y_f) = (-3/4, 3/4)$),  or to the centre of $D_1$ ($(x_f, y_f) = (-3/4, -3/4)$). 
The former case is denoted by $\vec{x}_f \in D_0$ and the latter by $\vec{x}_f \in D_1$.
This force added in the $D_1$ region and the corresponding vorticity forcing are shown in Figure~\ref{forcing}. 
Both forcing are localized.
The forcing amplitude is set to $a = \langle \psi \rangle_{D_j} \epsilon / \Delta t$, where $\Delta t$ is the simulation time step ($\Delta t = 10^{-3}$ for $Re \le 16000$) and $\langle \psi \rangle_{D_j}$ is the space-time average of the flow function without forcing in the region $D_j$.
We set the parameter to $\epsilon = 1.0\times 10^{-7}$ and check that values smaller than $\epsilon$ than this value give
the same result. 
The forcing duration is set to $T_f = 10$, which is about the same as the maximum delay time $10 \Delta t\times  \mathcal{E}\times \tau = 9$ used for embedding in Section \ref{sec:CCM}  (recall that COT representations are obtained at every 10 steps).
We also set $T_f$ to smaller values, e.g. $T_f = 0.5$, and check that the qualitative results do not change.

In the following, we observe $a^{-1}\Delta \omega$ to check the relation (\ref{delta_rel}). 
More specifically, we consider the average difference in the region $D_1$ by adding the forcing in the region $D_0$, which is denoted by
\begin{align}
  \left\langle a^
{-1}
  \Delta \omega \right\rangle_{D_0 \to D_1}
 &= \frac{1}{(\text{Area of $D_1$})}
     \left.\int_{D_1}  a^{-1}\Delta \omega(\vec{x}, t;\vec{x}_f, T_s+T_f)\right|_{\vec{x}_f \in D_0} ~{\rm d}{\vec{x}}.
\end{align}
Similarly, we define $\left\langle a^{-1}\Delta \omega \right\rangle_{D_1 \to D_0}$.
The average can be regarded as a suitable average of the linear response function $G_\omega(\vec{x}, t|\vec{x}', t')$ in view of equation (\ref{delta}).
We also calculate the difference of the streamfunction in the same manner
(we only consider the scalar quantities here). 

\begin{figure}
\centerline{\includegraphics[width=0.47\textwidth]{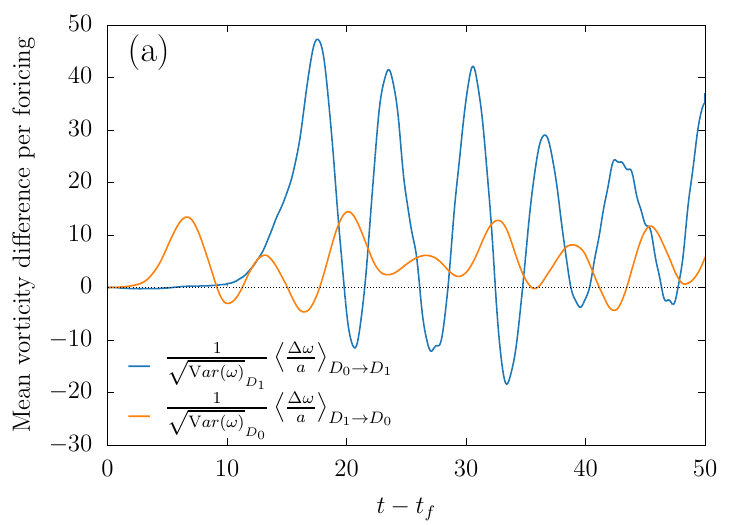}
\hspace*{0.6cm}
\includegraphics[width=0.47\textwidth]{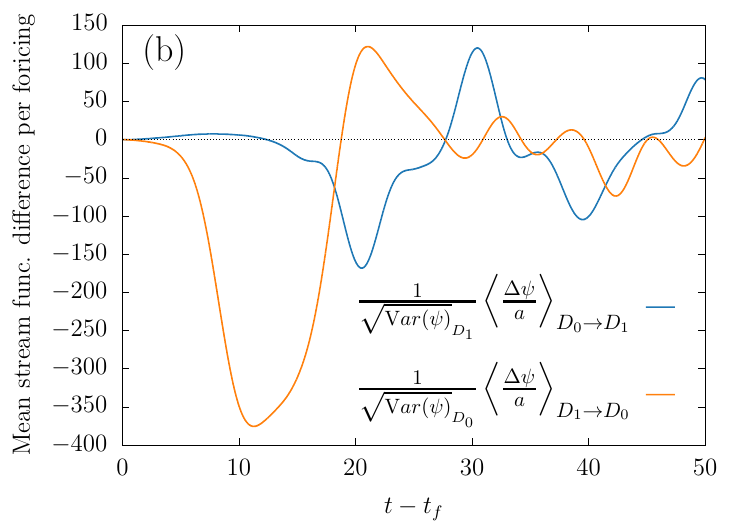}}  
\caption{\label{diff16k} (a) The mean vorticity differences per forcing amplitude, $a$, as a function of time. They are proxies of the linear response functions of the vorticity. Note that the differences are also normalised by the standard deviation of the (unforced) vorticity in the observed regions. Here, ${\rm Var}(\omega)_{D_j}$ means the variance of the vorticity averaged over in the region $D_j$ (the time average is taken as well). (b) The mean streamfunction differences. The Reynolds number is $Re=16000$ for both panels. The forcing is added here for $t - T_s \le T_f = 10$. The differences are observed from $t=T_s$ to $t=T_s + 5 T_f$ and the time average is taken over $100$ samples of the time extent $5T_f$.}
\end{figure}

\begin{figure}
\centerline{\includegraphics[width=0.47\textwidth]{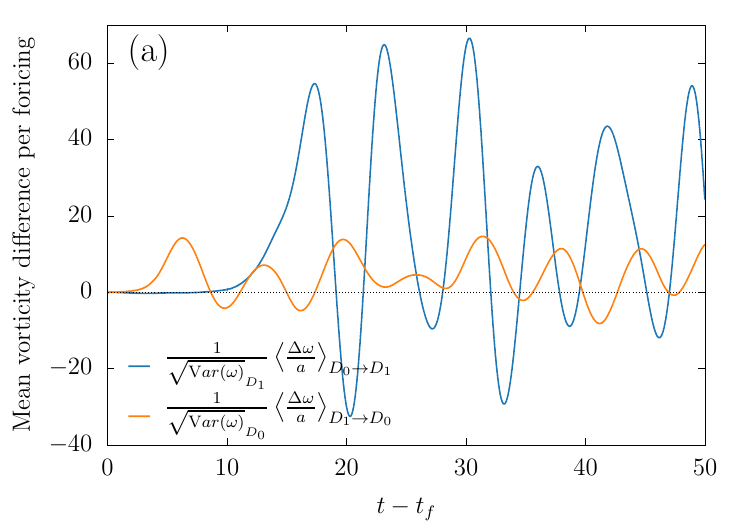}
\hspace*{0.6cm}
\includegraphics[width=0.47\textwidth]{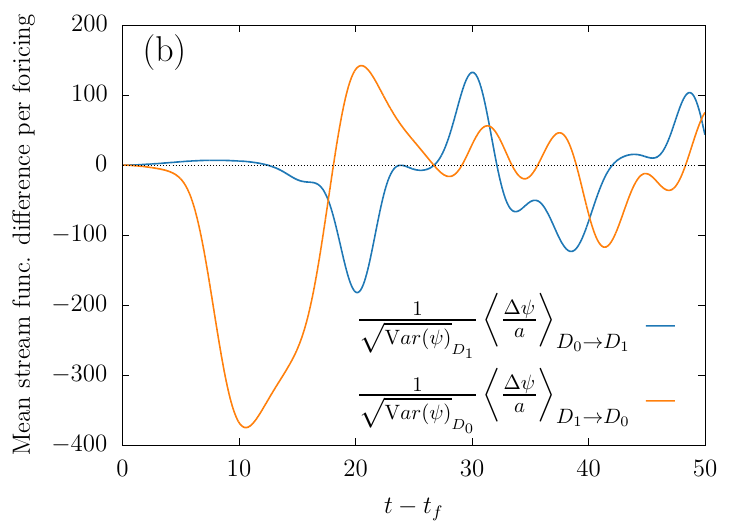}}  
\caption{\label{diff14k} Same as Figure~\ref{diff16k}, but for $Re=14000$. Those graphs are  similar to the ones for $Re=16000$.}
\end{figure}

For $Re = 16000$, we plot the averages of the differences in vorticity and stream function as a function of time in Figure~\ref{diff16k}.
The differences are normalised by the forcing amplitude and the standard deviations of the observed regions.
The amplitudes of the vorticity differences do not agree with our anticipation (\ref{Grel}) or (\ref{delta_rel}) as shown in Figure~\ref{diff16k}(a).
On the other hand, for the streamfunction, with which we identified the structures $c^1_+$ and $c^0_+$, the amplitude of the difference for $D_1 \to D_0$ is non-zero in $t \le 15$,  while the amplitude for $D_0 \to D_1$ is around zero, as observed in Figure~\ref{diff16k}(b).  However, in $t - t_f > 15$ the amplitudes become comparable. From this behaviour, we infer that $c^{1}_+$ has a causal influence on $c^{0}_+$ up to around $t - t_f = T_f$ \cite{bcv20} and that at later times no difference in causality is detected.
Therefore, it is not straightforward to conclude which corner eddy is causal in another one by observing the difference of the amplitudes.
For the lower Reynolds number $Re=14000$, where the flow is periodic in time, we plot the averaged differences as the linear response functions in Figure~\ref{diff14k}.
Their overall shapes remain similar to those for $Re=16000$.
Hence, we hardly find qualitative differences in linear responses across these Reynolds numbers.
This may be owing to the linearity of their governing equations and comparable P\'eclet numbers despite the underlying flow transitioning from periodic to chaotic states.
As a result, we find virtually no difference in sensitivity for periodic and chaotic evolutions. 
In this sense, the observational CCM approach based on TFDA can clarify the geometric causality between topological changes in flow patterns that the standard sensitivity analysis cannot capture.
The discrepancy between CCM and the linear response analyses can be attributed to their fundamentally different approaches to causality detection. CCM operates in the framework of state-space reconstruction, utilising Taken's embedding theorem to predict future states from historical time-series data. 
In the periodic regime illustrated in Figure~\ref{fig:Diagram_Re14000}(a), the COT representations exhibit highly regular patterns. Under these conditions, the system's future states can be predicted from any instantaneous state with equal accuracy, regardless of which corner's information is used. This makes causal relationships appear symmetric in the CCM analysis.
Linear response analysis, in contrast, provides a mechanistic view of causality by examining the system's response to controlled perturbations. This approach directly probes the spatio-temporal propagation of disturbances through the flow field. 
This methodological distinction highlights the complementary nature of observational (CCM) and interventional (linear response) approaches in understanding complex fluid systems. 

Finally, we discuss the relations of our work with recent studies on causality in fluid mechanics. 
In three-dimensional turbulent flows,  a standard way is to apply POD first and then study the causal relationships between the POD modes.  
In \cite{ld24}, the square duct flow was studied to uncover the cause of the secondary flow with the POD and the Granger causality analysis.  
In \cite{ms23}, the wake behind a circular cylinder mounted on a wall was analysed with the POD and the entropy-transfer method for causality.  
A time-series-based analysis for sub-grid energy fluxes with different grid scales in homogeneous and isotropic turbulence was performed in \cite{lda22} with the entropy-transfer method. 
An elaborate interventional study to date in this field is \cite{ld21} in which wall-bounded minimal-flow unit turbulence was analysed by numerically removing certain unstable modes to find the primary cause of the sustainment of turbulent fluctuations by the mean flow.  
In this paper, we studied the causality between the corner eddies represented by the COT symbols $c^0_+$ and $c^1_+$ as discrete modes in the lid-driven cavity flow.  
The difference from the previous work is that we used new modes and applied both approaches to causal analysis.
The advantages of the new modes, COTs, are the immediate interpretability and the robustness to noise owing to the topological approach. The COTs are not much influenced by noise, unlike the POD. 
The disadvantage is that it does not apply directly to three-dimensional flows, but we need to consider the projection of the flow onto an appropriate two-dimensional section, as we discussed in Summary.

\section{From Hamiltonian to COT representation}\label{sec:psiclone}
This appendix outlines the procedure for obtaining the COT representation from a given Hamiltonian. The process involves two main steps: first, the construction of a Reeb graph from the Hamiltonian scalar field and second, the conversion of this Reeb graph into its corresponding COT representation.

\subsection{Constructing the Reeb Graph from a Hamiltonian}

Let $H(x, y)$ be the Hamiltonian for a given 2D vector field. The flow streamlines are the level sets of this function. The Reeb graph is a skeletal representation that captures the topological evolution of these level sets.
Each point in a Reeb graph corresponds to a single connected component of a level-set of $H$. As the value of level $c$ is varied, the topology of the level set $H(x, y) = c$ changes: components may be created, destroyed, merged, or split. These topological events occur at the critical points of the Hamiltonian, which correspond to critical points in the flow field such as saddles and centres.
The critical points of $H$ form the vertices of the Reeb graph, and the edges represent the evolution of the level sets as $c$ varies. 
\begin{itemize}
\item The local minima (births) and maxima (deaths) of $H$ correspond to the leaf vertices, where connected components of the level sets appear or vanish.
\item Saddle points, where level-set components merge or split, become vertices of a degree greater than two.
\item Edges represent continuous changes in topology between critical levels and can be weighted by the difference in $H$-values between endpoints.
\end{itemize}

For discrete data (e.g., obtained from simulation or PIV), calculating analytically critical points is not feasible. 
Instead, ``birth'' and ``death'' of connected components in sublevel sets ($H\le c$) and super-level sets ($H\ge c$) are tracked by sweeping the Hamiltonian values from minimum to maximum. This generates two trees: a \emph{merge tree} for sublevel sets and a \emph{split tree} for super-level sets. These two trees are then merged to form the full Reeb graph.
Note that the Reeb graph is equivalent to the COT by construction~\cite{TFDA-UYS18}.

\subsection{Converting Reeb Graph to COT representation}
The conversion from a Reeb graph to a COT representation is performed recursively.
\begin{enumerate}
    \item When a threshold $\varepsilon>0$ is specified, the Reeb graph is simplified by removing edges with weights smaller than this threshold.
    \item The traversal starts from a designated root of the Reeb graph. For a flow in a bounded domain, a boundary component with the smallest value of $H$ is typically chosen as the root and given the symbol $\beta_{\emptyset}$.
    \item The algorithm traverses the graph from the root. At each vertex in the Reeb graph, it identifies the local structure on the basis of the vertex's degree and the nature of the corresponding critical point. 
    A COT symbol is assigned on the basis of the local structure.
    \begin{itemize}
        \item Degree-1 vertices map to elliptic centres ($\sigma_\pm$).
        \item Degree-3 vertices with self-connections map to $b$-type symbols ($b_{\pm\pm}$ or $b_{\pm\mp}$).
        \item Degree-($2+n$) vertices connecting distinct saddles on the boundary map to $c_\pm$, where $n$ is the number of local orbit structures which are attached to the boundary and are enclosed by $c_\pm$ together with a part of the boundary.
        \item The specific subscripts ($+$, $-$) are determined by the direction of the flow in the corresponding regions, which can be inferred from the Hamiltonian's values.
    \end{itemize}
    \item For each incident edge (in cyclic order), it recurs on the child subgraph, obtaining its COT symbols. These symbols become arguments of the parent COT symbol in the same cyclic order.
\end{enumerate}

The final COT string is constructed by concatenating the symbols assigned at each vertex, respecting the recursive structure of the Reeb graph.

\subsection*{Acknowledgements}
This research is supported by JST-Mirai Program Grant No. JPMJMI18G3 and JPMJMI22G1.

\bibliographystyle{abbrvnat}
\bibliography{jfm-lid}

\end{document}

%% file: physical_insight.tex
\subsection{Transition of flow complexity with topological states}\label{sec:vari_COT}
\begin{figure}
\begin{center}
\includegraphics[width=13cm]{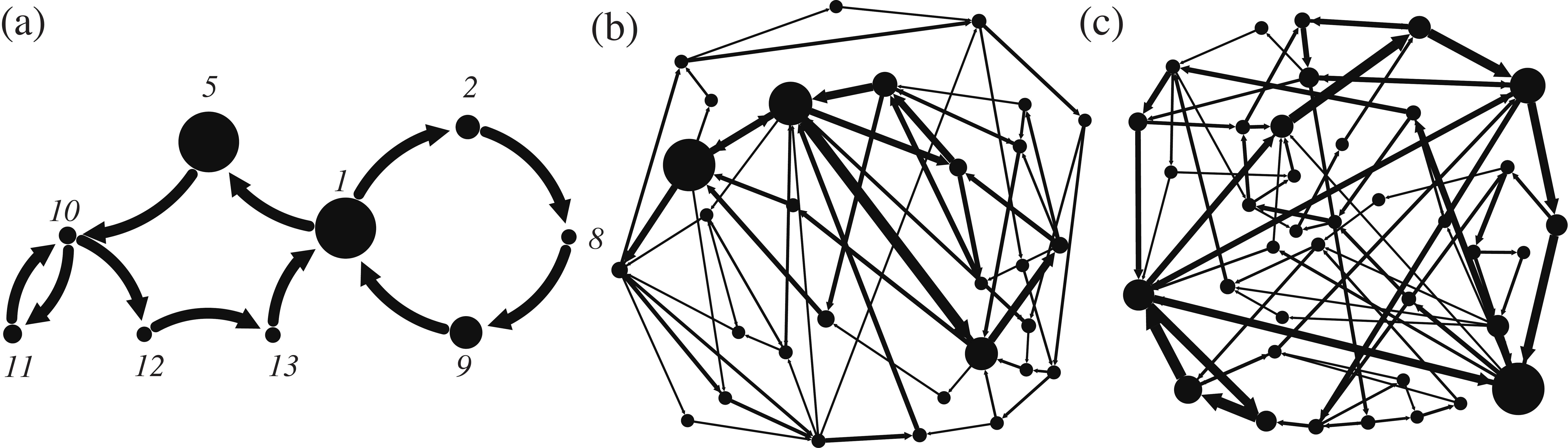}
\end{center}
  \caption{Transition diagrams for the lid-driven cavity flow for (a) $Re=15000$, (b) $Re=15500$, and (c) $Re=16000$.}
\label{fig:Diagram_transition_regime}
\end{figure}

We investigate the variation of topological states as a function of the Reynolds number.
In particular, we show that the increase in the number of topological states, or nodes in the transition diagram, has a critical behaviour around the transition from the periodic to the quasi-periodic state.
 We first visualise transition diagrams in the range of $15000 \le Re  \le 16000$, where the periodic behaviour changes to chaotic aperiodic behaviour according to \citet{shen91}.
Figure~\ref{fig:Diagram_transition_regime}(a) presents the transition diagram for $Re = 15000$.
The nodes are not labelled by COT representations but ID numbers instead due to insufficient display space.
The correspondence between ID numbers and COT representations is listed in Table~\ref{tab:cot_list}. 
These IDs function simply as numeric identifiers of topological states here and will be used consistently in subsequent sections.
The diagram shows that the flow is periodic and the evolution follows the sequence $1 \to 2 \to 8 \to 9 \to 1 \to 5 \to 10 \to 11 \to 10 \to 12 \to 13 \to 1$.
This constitutes a cycle of length $11$. 
However, since certain states share the same COT representation, the diagram appears to contain three distinct cycles. 
When the Reynolds number is $Re=15500$, the transition diagram becomes intricate, as shown in Figure~\ref{fig:Diagram_transition_regime}(b).
It indicates that the flow changes from periodic to aperiodic behaviour between $Re=15000$ and $15500$.
Figure~\ref{fig:Diagram_transition_regime}(c) presents the transition diagram for $Re=16000$, where a larger number of topological states are observed.

\begin{figure}
\centering
\includegraphics[width=12cm]{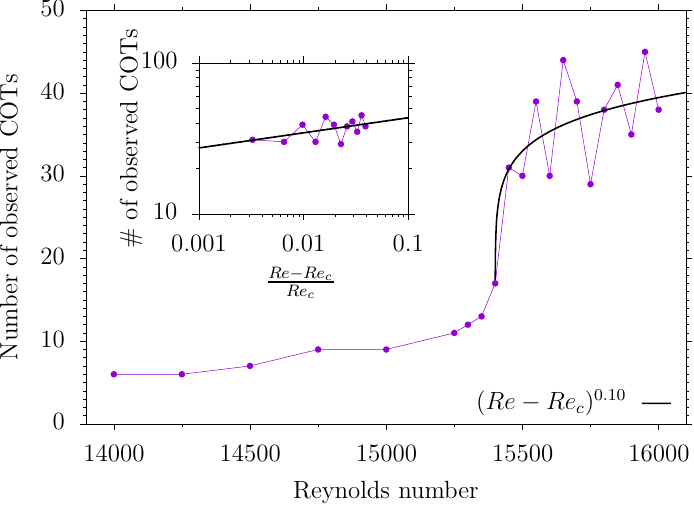} 
\caption{\label{fig:n_obs_cots} Number of observed topological states $n_{top}(Re)$ represented by COT representations as a function of Reynolds number.
Inset: a least-square fitting of the scaling behaviour around the critical Reynolds number which is here assumed to be $Re_c = 15400$. } 
\end{figure}

To quantify these regime changes in dynamics complexity, we simply count the number of distinct topological states $n_{top}(Re)$, as a function of the Reynolds number.
The resulting dependence is shown in Figure~\ref{fig:n_obs_cots}.
The quantity $n_{top}$ shows a sharp increase just below $Re=15500$, suggesting critical behaviour and serving as an order parameter for the change in the flow regime.
We fit this rapid growth with a power-law function, as illustrated in the inset of Figure~\ref{fig:n_obs_cots}.
With a critical Reynolds number $Re_c=15400$, we obtain a reasonable fit of the form
$n_{top}(Re) \sim (Re-Re_c)^{0.10}$.
It may also be of interest to determine the asymptotic scaling of
$n_{top}(Re)$ as $Re \to \infty$, but this is beyond the scope of this article.
Since a COT has a tree structure, as illustrated in Figure~\ref{fig:example-lid}, the number of possible representations can grow exponentially in this regime.

It should be noted that $n_{top}(Re)$ depends on the threshold $\varepsilon$ of TFDA, below which the spatial variations of the streamfunction are neglected, as described in Section~\ref{sec:3.1.1}. 
Hence, quantitative aspects of critical behaviour, such as the exponent value, also depend on the threshold $\varepsilon$. 
Nevertheless, we consider that the dependence becomes weak if the threshold is taken to be sufficiently small for a given Reynolds number.

\input{period} 

\subsection{Energy and enstrophy variations with topological states}\label{sec:energy_COT}
We consider the evolution of the kinetic energy $E(t)$ defined by (\ref{defE})
and that of the enstrophy defined by 
\begin{align}
 Q(t) = \int_{\Omega}  \frac{1}{2} (\omega(x, y, t))^2 ~{\mathrm d}x {\mathrm d}y.
 \label{defQ}
\end{align}
We here relate the topological states expressed by COT representations to the evolution of kinetic energy and enstrophy to identify their roles. 
In particular, we segment their time series according to topological states, which provides a new physical perspective of flow evolutions. 

\subsubsection{Periodic dynamics for $Re=14000$}
\begin{figure}
\centerline{%
\includegraphics[width=0.9\textwidth]{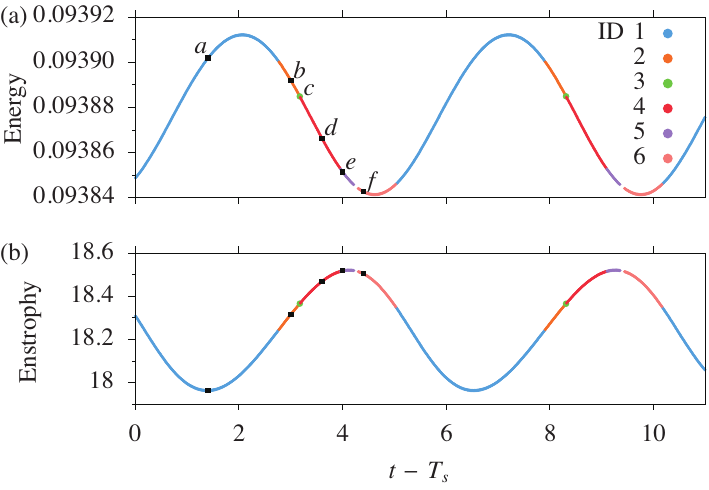}}
\caption{Variations of the flow for $Re=14000$; (a) energy and (b) enstrophy.
They are segmented with the six COT representations for $0 \le t-T_s \le 2.14\times \tau_p$, where the period is $\tau_p =5.15$.
The same colour on the graph means the flow patten belongs to a set of topologically equivalent class with the same COT representation.
The squares with labels $a$--$f$ on the evolution represents the representative topological states shown in Figure~\ref{fig:one_cycle_14000}(a)--(f).
The topological state with ID~$3$ appears for a short time.
Note that some data are missing between the ID~$5$ and $6$ (around $t-T_s=4.2$) due to an error of the COT computation with \texttt{psiclone}.}
\label{fig:eq14k_kcsv}
\end{figure}

Figure~\ref{fig:eq14k_kcsv} shows the temporal variations of energy and enstrophy
for $Re=14000$, which are periodic with the period $\tau_p=5.15$.
They are segmented with colours according to the six topological states. 
We assign ID numbers to the topological states with the same COT representation in order of appearance.
These IDs are the same as those introduced in Section~\ref{sec:vari_COT}.
As shown in Figure~\ref{fig:eq14k_kcsv}(a), while energy increases, the topological state is mainly ID $1$, which is  topologically equivalent to the flow pattern in Figure~\ref{fig:one_cycle_14000}(a).
Changes among the five remaining topological states occur while the energy decreases.
This implies that the five topological states with IDs $2$--$6$ shown in Figure~\ref{fig:one_cycle_14000}(b)--(f) are associated with relatively high energy dissipation.
Indeed, enstrophy of these states is higher than average, as indicated in Figure~\ref{fig:eq14k_kcsv}(b). 
Notice that enstrophy is proportional to the energy dissipation rate in the bulk of the lid-driven cavity flow. 
From this behaviour of the simple flow we infer that the states with relatively many closed streamline structures in the corners have larger enstrophy or larger energy dissipation.
This is in line with the physical intuition that the breakup of the corner vortices into smaller ones induces energy dissipation. 

\subsubsection{Periodic dynamics for $Re=15000$}
\begin{figure}
\centerline{\includegraphics[width=0.9\textwidth]{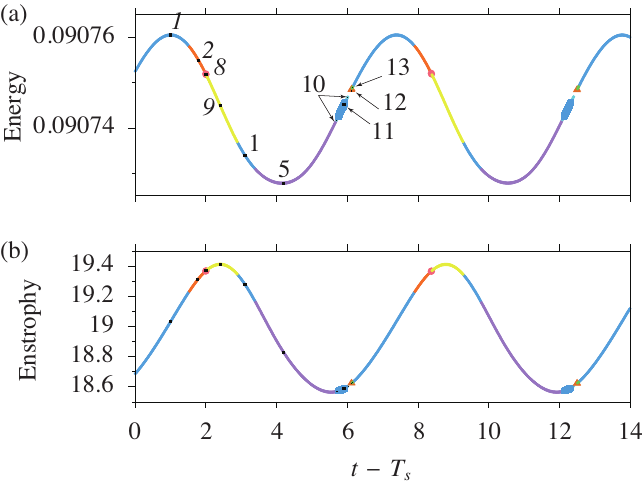}}
\caption{Variations of the flow for $Re=15000$; (a) energy and (b) enstrophy for $0\le t-T_s \le 2.2\times \tau_p$, where the estimated period is $\tau_p= 6.4$. 
The topological states with ID~$1$, $2$, $5$ are the same as those appeared for $Re=14000$, while those with ID $8$--$11$ are observed for the first time.
Data around $t-T_s = 6, 12$ is missing due to the error of the calculation of the COT.}
\label{fig:eq15k_kcsv}
\end{figure}

Figure~\ref{fig:eq15k_kcsv}(a) shows the energy variation which is color coded according to the topological states. 
The flow still exhibits a periodic behaviour in time with period $\tau_p=6.4$, but shows a qualitative change in the sequence of COT representations from the previous case. 
We observe nine topological states with IDs $1,2,5,8-13$ whose COT representations are listed in Table~\ref{tab:cot_list}. 
The topological states with ID~$1$, $2$ and $5$ have been present for $Re=14000$.
The qualitative change from the previous Reynolds number is that the state of ID~$1$ is observed in the energy-increasing and energy-decreasing phases. 
This splitting behaviour appears as the transition diagram with the separated cycles in Figure~\ref{fig:Diagram_transition_regime}(a). 
The central node connecting those cycles actually corresponds to the topological state with ID~$1$.

\begin{figure}
\centerline{%
\includegraphics[width=0.9\textwidth]{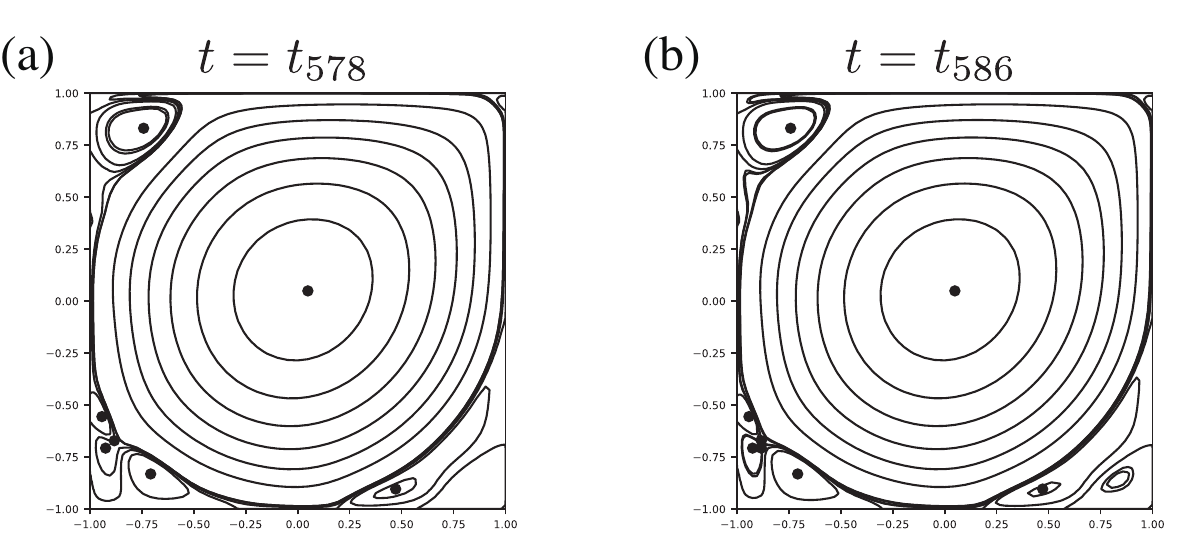}}
\caption{Streamline patterns at $t=t_{578}$ and $t=_{586}$ in the lid-driven cavity flow for $Re=15000$. They are identified as topological states with (a) ID $10$ and (b) ID $11$ respectively.}
\label{fig:top-states_15000}
\end{figure}

According to the transition diagram, the right cycle $1\rightarrow 2 \rightarrow 8 \rightarrow 9 \rightarrow 1$ corresponds to the energy-decreasing phase.
This route corresponds to the local change in the flow topology in the top left corner, since we observe the change in the $c_+^0$ sequences of the COT representations in Table~\ref{tab:cot_list}.
Hence, the topological change in the eddy structures in the top left corner serves to reduce the energy.
In contrast, Figure~\ref{fig:eq15k_kcsv}(a) also shows that the energy-increasing phase takes the route $1 \rightarrow 5 \rightarrow 10 \rightarrow 11 \rightarrow 10 \rightarrow 12 \rightarrow 13 \rightarrow 1$, which appears as the left cycles of the transition diagram in Figure~\ref{fig:Diagram_transition_regime}(a).
We simply see  the change in the sequence of $c_+^1$ for the topological states with ID $5$, $12$ and $13$ in Table~\ref{tab:cot_list}.
However, the COT representations for the topological states with ID $10$ and $11$ look different; therefore, we examine these flow patterns closely.
Figure~\ref{fig:top-states_15000}(a) is the streamline pattern at $t=t_{578}$ for $Re=15000$, which is identified as the topological state with ID~$10$.
Its COT representation contains the COT symbol $b_{--}\{ \sigma_-, \sigma_-\}$, indicating the existence of a figure-eight structure.
This structure appears as a small self-connected saddle connection between the two boundary saddle connections of $c_+^1$ in the bottom left corner, as schematically shown in Figure~\ref{fig:h-unstable}(c).
The streamline pattern at $t=t_{586}$ in Figure~\ref{fig:top-states_15000}(b) is in the topological state with ID~$11$.
It has a small structure of nested saddle connections, which is represented by $b_{-+}(\sigma_-, b_{-+}(\sigma_-,\sigma_+))$ in the COT representation, between the two boundary saddle connections of $c_+^1$ in the bottom left corner.
Hence, we conclude that the change in the eddy structures in the bottom left corner gives rise to the energy-increasing phase.

Another notable feature of Figure~\ref{fig:eq15k_kcsv} is that the duration of the topological state of ID~$5$ becomes longer than in the previous $Re=14000$ case.
This is shown as a large node in the left cycle of the transition diagram in Figure~\ref{fig:Diagram_transition_regime}(a).

\begin{table}
\centering
\caption{COT representations and IDs used in Section~\ref{sec:vari_COT}.}
\label{tab:cot_list}
\begin{adjustbox}{scale=0.9}
\begin{tabular}{ccp{9cm}}
\hline
\textbf{ID} & \textbf{First appearance} & \textbf{COT representation} \\ \hline
1 & $Re=14000$ & $\beta_{\emptyset-}(\sigma_{-}, \{c_{+}^{0}(\sigma_{+}) \cdot c_{+}^{1}(\sigma_{+}) \cdot c_{+}^{2}(\sigma_{+})\})$\\
2 & $Re=14000$ & $\beta_{\emptyset-}(\sigma_-, \{c_+^0(\sigma_+) \cdot c_+^0(\sigma_+) \cdot  c_+^1(\sigma_+) \cdot c_+^2(\sigma_+)\})$ \\
5 & $Re=14000$ & $\beta_{\emptyset-}(\sigma_{-}, \{c_{+}^{0}(\sigma_{+}) \cdot c_{+}^{1}(\sigma_{+}) \cdot c_{+}^{1}(\sigma_{+}) \cdot c_{+}^{2}(\sigma_{+})\})$ \\ 
8 & $Re=15000$ & $\beta_{\emptyset-}(\sigma_{-}, \{ c_{+}^{1}(\sigma_{+}) \cdot c_{+}^{2}(\sigma_{+})\})$ \\ 
9 & $Re=15000$ & $\beta_{\emptyset-}(\sigma_-, \{c_+^0(b_{++}\{\sigma_+, \sigma_+\}) \cdot c_+^1(\sigma_+) \cdot c_+^2(\sigma_+) \})$ \\ 
10 & $Re=15000$ & $\beta_{\emptyset-}(b_{--}\{ \sigma_-, \sigma_-\}, \{ c_{+}^{0}(\sigma_{+}) \cdot c_{+}^{1}(\sigma_{+}) \cdot c_{+}^{1}(\sigma_{+}) \cdot c_{+}^{2}(\sigma_{+})\})$ \\ 
11 & $Re=15000$ & $\beta_{\emptyset-}(b_{-+}( \sigma_-, b_{+-}(\sigma_+,\sigma_-)), \{ c_{+}^{0}(\sigma_{+}) \cdot c_{+}^{1}(\sigma_{+}) \cdot c_{+}^{1}(\sigma_{+}) \cdot c_{+}^{2}(\sigma_{+})\})$ \\ 
12 & $Re=15000$ & $\beta_{\emptyset-}(\sigma_-, \{c_+^0(\sigma_+)\cdot c_+^1(b_{++}\{\sigma_+, \sigma_+\},c_-(\sigma_-))\cdot c_+^2(\sigma_+)\})$ \\
13 & $Re=15000$ & $\beta_{\emptyset-}(\sigma_-, \{c_+^0(\sigma_+)\cdot c_+^1(\sigma_+,c_-(\sigma_-))\cdot c_+^2(\sigma_+)\})$ \\
\hline
33 & $Re=15500$ & $\beta_{\emptyset-}(\sigma_-, \{c_+^0(\sigma_+)\cdot c_+^1(\sigma_+)\cdot c_+^2(b_{++}\{\sigma_+, \sigma_+\})\})$ \\ 
58 & $Re=15500$ & $\beta_{\emptyset-}(\sigma_-, \{c_+^0(\sigma_+)\cdot c_+^1(\sigma_+)\cdot c_+^2(\sigma_+, c_-(\sigma_-))\})$ \\ \hline
60 & $Re=16000$ & $\beta_{\emptyset-}(\sigma_-, \{c_+^0(\sigma_+)\cdot c_+^1(\sigma_+)\cdot c_+^1(\sigma_+)\cdot c_+^2(b_{++}\{\sigma_+, \sigma_+\})\})$ \\
61 & $Re=16000$ & $\beta_{\emptyset-}(\sigma_-, \{c_+^0(\sigma_+)\cdot c_+^1(b_{++}\{\sigma_+, \sigma_+\})\cdot c_+^2(b_{++}\{\sigma_+, \sigma_+\})\})$  \\ \hline
\end{tabular}
\end{adjustbox}
\end{table}

\subsubsection{Quasi-periodic dynamics for $Re=15500$}
\begin{figure}
\centerline{%
\includegraphics[width=0.9\textwidth]{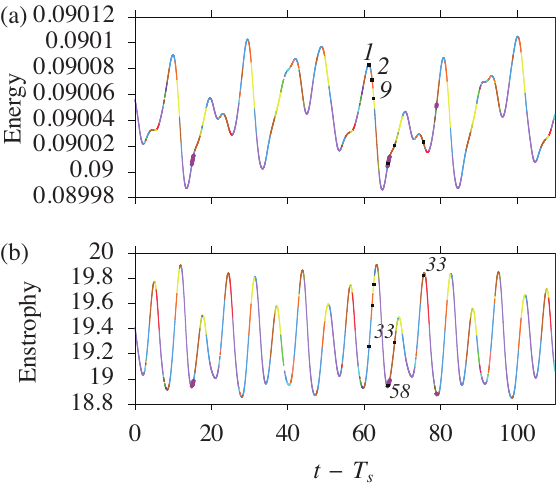}}
\caption{Same as Figure~\ref{fig:eq14k_kcsv} but for $Re=15500$ in $0 < t - T_s \le 110$. 
The COT representations for the topological states are listed in Table~\ref{tab:cot_list}.
(a) Variation of the energy.
We observe the route $1\rightarrow 2 \rightarrow 9$ around the energy maxima.
(b) Variation of the enstrophy.
The topological state with ID~$33$ that is newly observed for $Re=15500$ appears several place with the high rate of occurrence.
The topological state with ID~$58$ appears around the local enstrophy minima with a relatively high rate of occurrence.
Some data are missing due to the error of the calculation of the COT}
\label{fig:eq15.5k_kcsv}
\end{figure}

For $Re=15500$, the power spectrum of the energy time series indicates that the flow is considered quasi-periodic, as shown in Figure~\ref{fig:spc}(b).
We show the temporal variations of energy and enstrophy in Figure~\ref{fig:eq15.5k_kcsv}. 
We then observe thirty different topological states and complex transitions between them according to the transition diagram in Figure~\ref{fig:Diagram_transition_regime}(b).
However, in Figure~\ref{fig:eq15.5k_kcsv}(a), the route $1 \rightarrow 2 \rightarrow 9$ of the topological states, which has been observed for $Re=15000$, frequently appears around the local energy maxima.
Also, around the local energy minima, we observe the topological state with ID~$5$ that is the same as in the periodic regimes. 

Since we have many topological states in the quasi-periodic regime, we focus on the five states that are most frequently observed in the time series, which are visualised as large nodes of the transition diagram in Figure~\ref{fig:Diagram_transition_regime}(b).
Four of them are the topological states that have appeared in periodic regimes; the first place is ID~$5$ ($0.30$), the second place is ID~$1$ ($0.24$), the fourth place is ID~$9$ ($0.091$), and the fifth place is ID~$2$ ($0.036$).
The numbers in parentheses are the rate of occurrence, which will be discussed in detail later.
The third most frequently observed topological state (ID~$33$), which newly appears in this quasi-periodic regime, has the rate of occurrence $0.15$.
Its COT representation is given by $\beta_{\emptyset-}(\sigma_-, \{c_+^0(\sigma_+)\cdot c_+^1(\sigma_+)\cdot c_+^2(b_{++}\{\sigma_+, \sigma_+\})\})$ as shown in Table~\ref{tab:cot_list}.
This indicates that the change of the local topological structure in the bottom right corner, which has not been observed in periodic regimes, contributes to the increase in topological states.
This topological state appears not only around the local maxima of enstrophy, but also around a middle value of enstrophy, leading to a splitting behaviour as shown in Figure~\ref{fig:eq15.5k_kcsv}(b).

Another newly observed topological state (ID~$58$) with a relatively longer duration $0.012$ in this quasi-periodic regime has the COT representation $\beta_{\emptyset-}(\sigma_-, \{c_+^0(\sigma_+)\cdot c_+^1(\sigma_+)\cdot c_+^2(\sigma_+, c_-(\sigma_-))\})$.
Since it appears mainly as minimum enstrophy, it can be compared with a similar enstrophy-minimum topological state (ID~$10$) for $Re=15000$, which remains observed for $Re=15500$, but its occurrence rate decreases from $0.0306$ ($Re=15000$) to $0.0037$ ($Re=15500$). 
The difference in the COT representations indicates that the local streamline structures in the corners at the bottom are different; 
the local eddy structure is present in the bottom right corner for this topological state, while it is in the bottom left corner for the ID~$10$ state.
We infer that facilitation of eddy structures in the bottom right corner ($c_+^2$) is preferred in the enstrophy minima over having more structures in the bottom left corner for $Re=15500$.

\subsubsection{Chaotic dynamics for $Re=16000$}
\begin{figure}
%
\centerline{%
\includegraphics[width=0.9\textwidth]{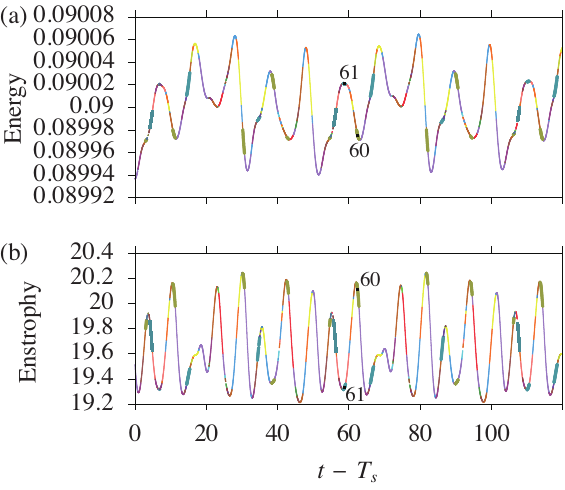}}
\caption{Same as Figure~\ref{fig:eq14k_kcsv} but for $Re=16000$ in $0 < t - T_s \le 120$. 
(a) Variation of the energy.
Points for IDs~$60$ and $61$ are enlarged to distinguish them from others. 
(b) Variation of the enstrophy.
Some data are missing due to the error of the calculation of the COT.}
\label{fig:eq16k_kcsv}
\end{figure}

The final example is the case of $Re=16000$ where the flow is considered chaotic. 
We observe thirty eight topological states in the time series, as shown in Figure~\ref{fig:Diagram_transition_regime}(c).
In Figure~\ref{fig:eq16k_kcsv}(a), we observe that some features are in common with those of $Re=15500$. 
That is, the route of topological states ID~$1 \rightarrow 2 \rightarrow 9$ is still intermittently observed around the local energy maxima, for instance around $t-T_s \approx 80$.
In addition, the topological state with ID~$5$ is still observed around most of the local energy minima.
The top-five topological states that occur most frequently are ID~$5$ ($0.22$), $33$ ($0.13$), $1$ ($0.10$), $9$ ($0.085$) and $58$ ($0.056$), where the numbers in parentheses are the rate of occurrence.
It is important to note that they have already been observed in the periodic and quasi-periodic regimes, and the occurrence rate of the ID~$58$ state exceeds that of the ID~$2$ state.
The statistical properties of the occurrence rate of topological states play an important role in understanding chaotic dynamics and are therefore examined in detail in the next subsection.

As shown in Figure~\ref{fig:Diagram_transition_regime}(c), there are a large number of topological states, and the transitions between them become complicated and uneven in the transition diagram.
Hence, it is generally difficult to extract a particular route of changes in the COT representation.
However, we pick the topological states of ID~$60$ ($0.048$) and $61$ ($0.041$) with a higher occurrence rate than $0.01$, whose COT representations are provided in Table~\ref{tab:cot_list}.
They indicate that the difference in the eddy structure is observed in the bottom left corner.
That is, the ID~$60$ state has two isolated eddy regions, while the ID~$61$ state has a figure-eight-shaped streamline structure there. 
The variation in enstrophy in Figure~\ref{fig:eq16k_kcsv}(b) indicates two features on the ID $60$ and $61$ states.
Firstly, the topological state with ID~$60$ appears around extreme enstrophy values, more specifically, primarily around local maxima and occasionally around local minima of enstrophy.
Secondly, the state with ID~$61$ occurs after some local minima.
This observation implies that a figure-eight streamline on the bottom left corner is not favourable for having the enstrophy local maxima. 

\subsection{The rate of occurrence across Reynolds numbers}\label{sec:occurrence}
We have assigned identification numbers (ID) to the topological states with the COT representation in order of appearance in the time series. 
They are useful for identifying topological states with the same COT representations across Reynolds numbers.
That is, regardless of Reynolds numbers, the topological state with ID~$1$ uniquely refers to the same topologically equivalent flow pattern.

\begin{figure}
\centering
\includegraphics[width=0.8\textwidth]{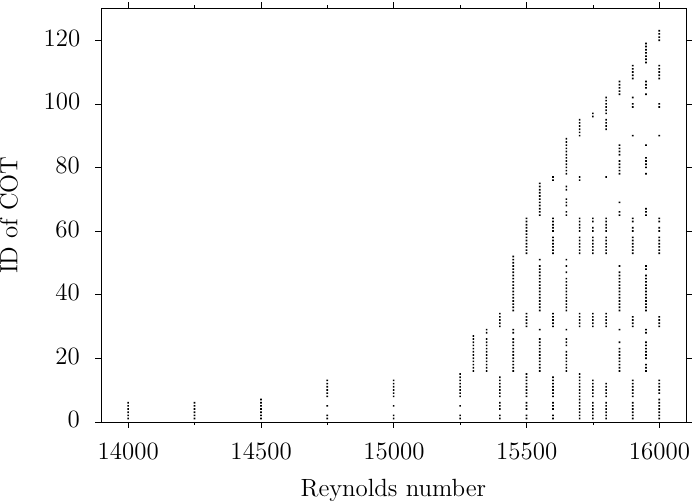}
\caption{Topological states observed at different Reynolds numbers.
Each square dot indicates that the topological state with the ID appears at the Reynolds number. 
The maximum ID number at $Re=16000$ is $123$, which depends on the sampling rate and the Reynolds numbers.} 
\label{fig:cot_id} 
\end{figure}

Figure~\ref{fig:cot_id} shows which IDs of topological states or COT representations are observed at each Reynolds number.
We have previously plotted the total of these topological states as a function of the Reynolds number in Figure~\ref{fig:n_obs_cots}. 
Now, the ID diagram conveys different information about how the complexity of the flow increases as the Reynolds number increases. 
An envelope of the diagram indicates that the distribution of the topological states takes three different regimes: 
(i) $14000 \le Re \le 14500$, (ii) $14750 \le Re \le 15250$, and (iii) $15300 \le Re$, although the envelope depends on how we sample both the time series and the Reynolds numbers.

The regimes (i) and (ii) correspond to periodic flow evolutions, while the regime (iii) may correspond to quasi-periodic evolutions.
The change from (i) to (ii) is relatively minor compared to the change from (ii) to (iii).
It is also difficult to detect the first change with, for example, the power-spectrum analysis of the energy.
This first change may correspond to the emergence of the three cycles of the transition diagram in Figure~\ref{fig:Diagram_transition_regime}(a).
The second change is accompanied by the noticeable increase in the number of observed topological states.
In regime (iii), we observe a clustering of the IDs at each Reynolds number, which takes the form of vertical segments of IDs.
For example, the segment of topological states from ID~$16$ to $27$ first appears at $Re=15300$, disappears at $Re=15400$, and reappears at $Re=15450$ with a small change.
The appearance and disappearance occur alternately.  
This alternating behaviour gives rise to oscillations in the number of observed topological states for $Re > 15400$ as we see in Figure~\ref{fig:n_obs_cots}, although we have not yet find its physical interpretation.
At the same time, new topological states emerge and are piled on top segments in each Reynolds number in Figure~\ref{fig:cot_id}.  
The figure can visually describe how the complexity of the flow increases in the quasi-periodic regime. 
The peculiar pattern in Figure~\ref{fig:cot_id} is reminiscent of windows of various periodic motions in the bifurcation diagram of the logistic map beyond the period doubling cascade \citep{Ott_2002}, although the lid-driven cavity flow is not considered so chaotic.
Of course, more fine structures in the segment pattern may be found, if we sample Reynolds numbers more frequently in the diagram.
We do not find a different signature for $Re=16000$, which possibly differentiates the chaotic state from the quasi-period state in the diagram, contrary to our argument based on the power spectrum of the energy in Section~\ref{subsec:2.1}.

\begin{figure}
\centering 
\includegraphics[width=0.7\textwidth]{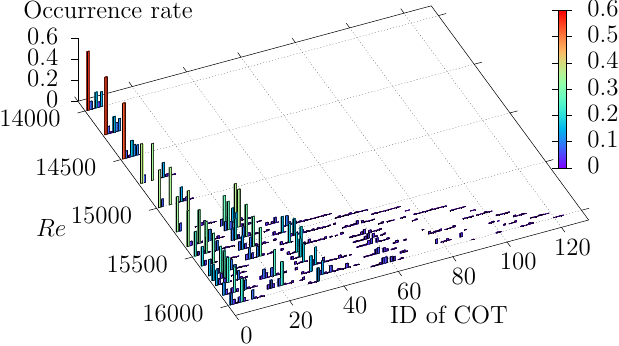}
\caption{Occurrence rate of topological states or COTs at each Reynolds number in $20000$ snapshots for $0 < t - T_s \le 200$.}\label{fig:occurence_rate}
\end{figure}

In Section \ref{sec:energy_COT}, for $Re=15500$ and $16000$, we focus on some topological states that persist for a long time. 
The persistence of a topological state has been measured by the rate of occurrence, which is the number of snapshots identified as this state divided by the total number of snapshots.
We plot this rate as a function of the Reynolds number and the ID in Figure~\ref{fig:occurence_rate}, which illustrates how newly born topological states participate statistically as the flow changes from periodic to quasi-periodic and chaotic states. 
At the same time, it shows that the topological states already existing at low Reynolds numbers keep appearing in the flow evolution at higher Reynolds numbers. 
Topological states that exist at low Reynolds numbers, say, with IDs less than $10$, keep a relatively high rate of occurrence at higher Reynolds numbers. 
This point is interesting, since it is argued in general that time-averaged flow patterns of fully-developed turbulent flow resemble those at much lower Reynolds numbers \citep{da15}.
Here, we see that this argument follows from the relatively high rates of occurrence of the small Reynolds number topological states even at larger Reynolds numbers.
Nevertheless, the occurrence rate of the topological states with small IDs slowly decreases as a function of Reynolds number.
In contrast, at $Re=15300$ the states with IDs around $20$ are newly born. 
Their occurrence rate is high from the beginning since low ID states are not observed at $Re=15300$ and gradually decreases as the Reynolds number increases.
On the other hand, the topological states around ID~$35$ or ID~$60$ have a lower occurrence rate at birth. 
Then their rates increase gradually and turn to decrease as the Reynolds number increases.
Probably, this trend continues for newly born topological states at larger Reynolds numbers. 

%% file: period.tex
\subsection{Period estimation}\label{sec:period}
One of the physical quantities that characterizes periodic motion is its period.
The most common approach to determining the period is to apply a Fast Fourier Transform (FFT) to a time series of a physical observable such as kinetic energy and to analyse the resulting power spectrum.
The period is estimated from the peak frequency of the power spectral density, as we did in Section \ref{subsec:2.1} for $Re=14000$. 
In addition to such conventional physical observables, we here demonstrate that the period can also be estimated from time series of COT representations.
More precisely, by computing the COT representation for each snapshot and arranging these descriptors in temporal order, we obtain a topology-based time series.
Accordingly, we estimate characteristic periods from two distinct types of time series:
(i) a continuous physical observable, the kinetic energy, and
(ii) discrete signals derived from TFDA, given by the temporal evolution of COT descriptors.
This framework enables a direct comparison between physics-based and topology-based descriptions of the dynamics.
In particular, we will show that the estimated periods from the energy and the COT representation are quantitatively consistent, implying that the COT representation indeed reflects the essential physical information of the flow.

In addition to the spectral method (FFT), we employ two complementary period estimators.
The first is based on the autocorrelation function (ACF); see \citet[Chapter 2]{proakis2007digital}.
Autocorrelation measures the similarity between a signal and a time-shifted copy of itself as a function of the time lag; periodic signals produce peaks at lags corresponding to integer multiples of the period.
For continuous-valued signals, we first rescale the data so that they have zero mean and unit variance (standard score, i.e., Z-score normalisation).
For discrete COT-based signals, the symbolic descriptors are mapped to numerical values via a hash-based projection so that correlation operations can be performed.\footnote{Because this projection depends on the specific hash mapping, the resulting correlations can vary; we retain this result here for reference.}
Dominant periods are then identified by detecting prominent peaks in the autocorrelation function.

The second estimator is based on Dynamic Time Warping (DTW) proposed by~\citet{saoke78}.
DTW evaluates the similarity between two temporal segments, while allowing for local stretching or compression along the time direction.
To estimate candidate periods, the time series is divided into adjacent segments of a given length, each regarded as a cycle.
For each candidate period, we compute the DTW distance between neighbouring segments in a constrained warping window (2 seconds).
The mean warping distance over multiple shifted segmentations is then calculated, providing a quantitative measure of how consistently the temporal pattern repeats.
A true period is expected to yield a small average warping distance, reflecting strong cycle-to-cycle similarity.

Table~\ref{tab:periods} summarises the estimated dominant periods for the lid-driven cavity flow, obtained from both kinetic energy and COT-based time series using the FFT, ACF and DTW estimators.
Up to $Re = 15250$, the estimated periods are consistent between observables and estimation methods. In this regime, the COT representation can be effectively regarded as a label identifying distinct phases within a single period. 
However, for $Re \ge 15300$, substantial discrepancies emerge between estimators and between signal types, indicating that the notion of a dominant period is no longer meaningful in this regime. 
A possible physical interpretation of this discrepancy is that the flow at $Re=15300$ is not strictly periodic.
By applying the two estimation methods to two observables, the energy and the COT representation, we infer that the quasi-periodic regime observed at $Re=15500$ extends down to $Re=15300$.
Consequently, the transition from the periodic state to the quasi-periodic state is likely to occur within the interval $15250 < Re < 15300$.

\begin{table} \centering 
\caption{Estimated dominant periods. For each Reynolds number, we report the period obtained from the kinetic energy time series and from a COT-based time series, using the ACF and DTW estimators. The periods estimated with the power spectral density of the energy, denoted by FFT, are listed up to $Re=15000$.} \label{tab:periods} 
\begin{adjustbox}{scale=0.9}
\begin{tabular}{rcccccc} \hline $Re$ & Energy (FFT) & Energy (ACF) & Energy (DTW) & COT (ACF) & COT (DTW) \\ 
\hline 
14000 & 5.15 & 5.132 & 5.090  & 5.142 & 5.090 \\
14250 & 5.13 & 5.147 & 5.137  & 5.147 & 5.076 \\
14500 &  5.13 & 5.159 & 5.141  & 5.169 & 5.110 \\
14750 & 6.45 & 6.360 & 6.373  & 6.370 & 6.191 \\
15000 & 6.45 & 6.385 & 6.406  & 6.385 & 6.193 \\
15250 & -- & 6.397 & 6.426  & 6.386 & 6.244 \\
15300 & -- & 19.207 & 6.460  & 19.187 & 6.491 \\
15350 & -- & 19.298 & 6.761  & 19.156 & 6.367 \\
15400 & -- & 19.579 & 24.812  & 19.398 & 24.812 \\
15450 & -- & 19.718 & 24.758  & 19.406 & 12.907 \\
15500 & -- & 19.843 & 24.847  & 19.401 & 12.893 \\
15550 & -- & 19.872 & 24.753  & 19.449 & 24.753 \\
15600 & -- & 20.002 & 11.045  & 19.438 & 13.290 \\
15650 & -- & 20.013 & 10.634  & 19.560 & 13.195 \\
15700 & -- & 10.412 & 10.930  & 19.533 & 24.857 \\
15750 & -- & 20.220 & 10.108  & 19.476 & 24.773 \\
15800 & -- & 20.184 & 10.380  & 19.490 & 12.552 \\
15850 & -- & 10.348 & 10.880  & 19.520 & 24.743 \\
15900 & -- & 10.289 & 10.452  & 19.614 & 24.729 \\
15950 & -- & 20.342 & 9.983  & 19.607 & 24.837 \\
16000 & -- & 10.323 & 10.409 & 19.779 & 24.916 \\
\hline \end{tabular} 
\end{adjustbox}
\end{table}

%% file: CCM.tex
This section investigates causal relations among the topological changes in flow patterns that occur in the corners of the cavity.
Using COT representations, the evolution of the flow field is described as a sequence of discrete topological states.
In particular, the structural changes of the recirculating regions in each corner of the cavity appear as a sequence of topological events. 
This sequence of events naturally raises the question of how the topological changes occurring at different corners influence each other during flow evolution.
For instance, when a recirculating region appears in the top left corner $c_+^0$, it is not immediately clear whether this change is induced directly by the lid-driven shear near the top boundary, or whether it arises indirectly through the transport of vorticity generated by the recirculating region in the bottom left corner $c_+^1$ through large-scale circulation inside the cavity.
Addressing such questions requires a framework capable of quantifying causal influence between sequences of topological events extracted from the flow.
In this study, therefore, we analyse causal relations between the sequences of topological states obtained from the COT representation, with particular attention to the formation and disappearance of corner recirculating regions represented by $c^0_+$ and $c^1_+$.

From a mathematical viewpoint, the incompressible Navier--Stokes equations determine the flow field uniquely under given boundary conditions, so the notion of causality may appear unnecessary.
However, causal analysis has been employed in a fluid mechanical problem to rigorously examine a physical picture in order to obtain a better understanding of the underlying physics; see, e.g.,  causal analyses of the self-sustainment mechanism in wall turbulence by \cite{ld20, ld21}. This approach is considered an insightful approach to develop better turbulence modelling and control.
Although the pressure field adjusts instantaneously in the incompressible formulation, the transport of momentum and vorticity, which governs the formation and disappearance of corner recirculating regions, occurs through advection and diffusion processes over finite time scales.
Consequently, changes in the local flow structure at one corner can physically influence the evolution of structures at other corners with a finite delay.
In this manner, this causality analysis enables us to develop a justifiable interpretation of intricate interactions among flow structures.

Here we employ Convergent Cross Mapping (CCM) developed by \cite{CCM} to measure causality among flow pattern changes.
This causal inference can be classified as observational, following the classification in \citet{bcv20}.
This approach investigates whether knowledge of one variable is useful in predicting future values of another. 
This framework is well suited to analysing sequences of COT representations.
An interventional analysis based on continuous quantities, which is commonly employed in fluid mechanics because of their direct relevance to active flow control, is presented in Appendix \ref{sec:Appendix}.

\begin{figure}
\begin{center}
\includegraphics[bb=0 0 1008 360, width=0.9\textwidth]{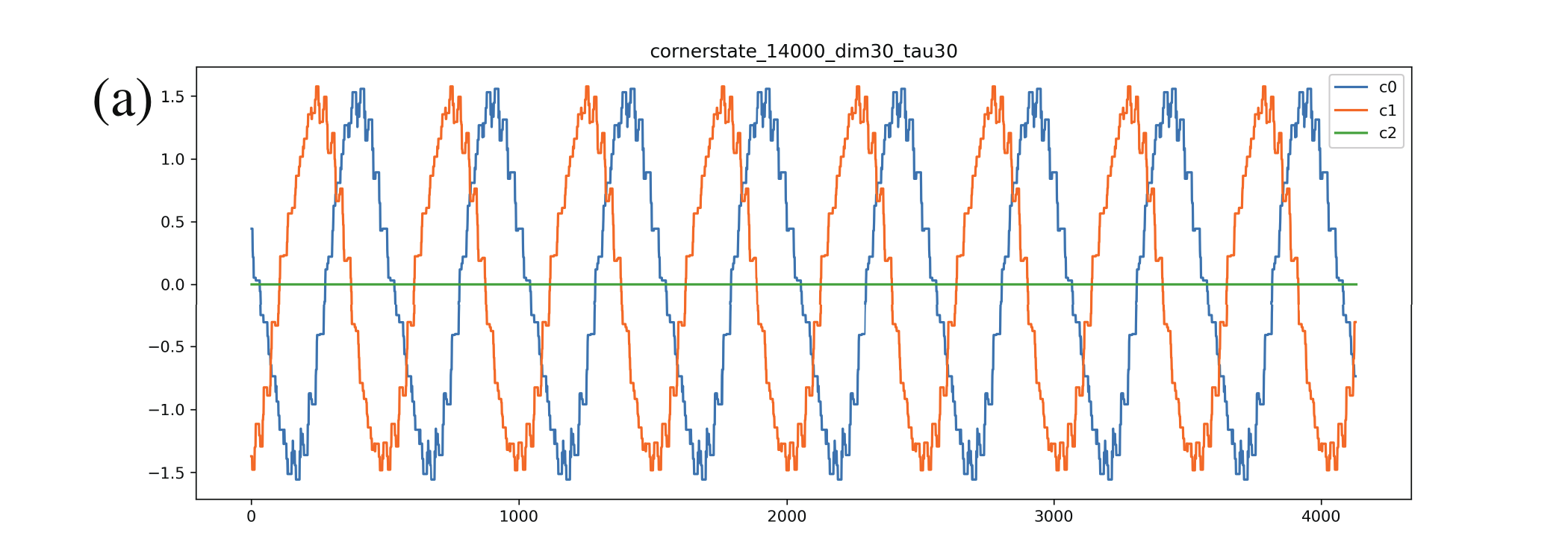}
\includegraphics[bb=0 0 1008 360, width=0.9\textwidth]{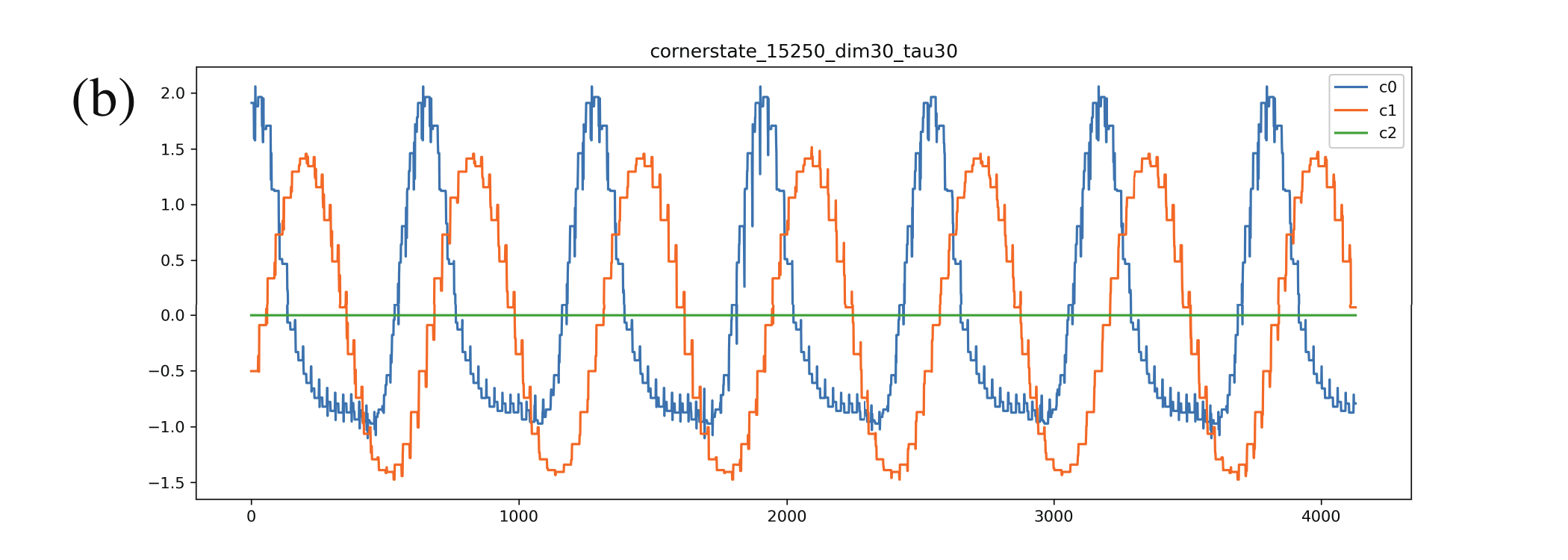}
\includegraphics[bb=0 0 1008 360, width=0.9\textwidth]{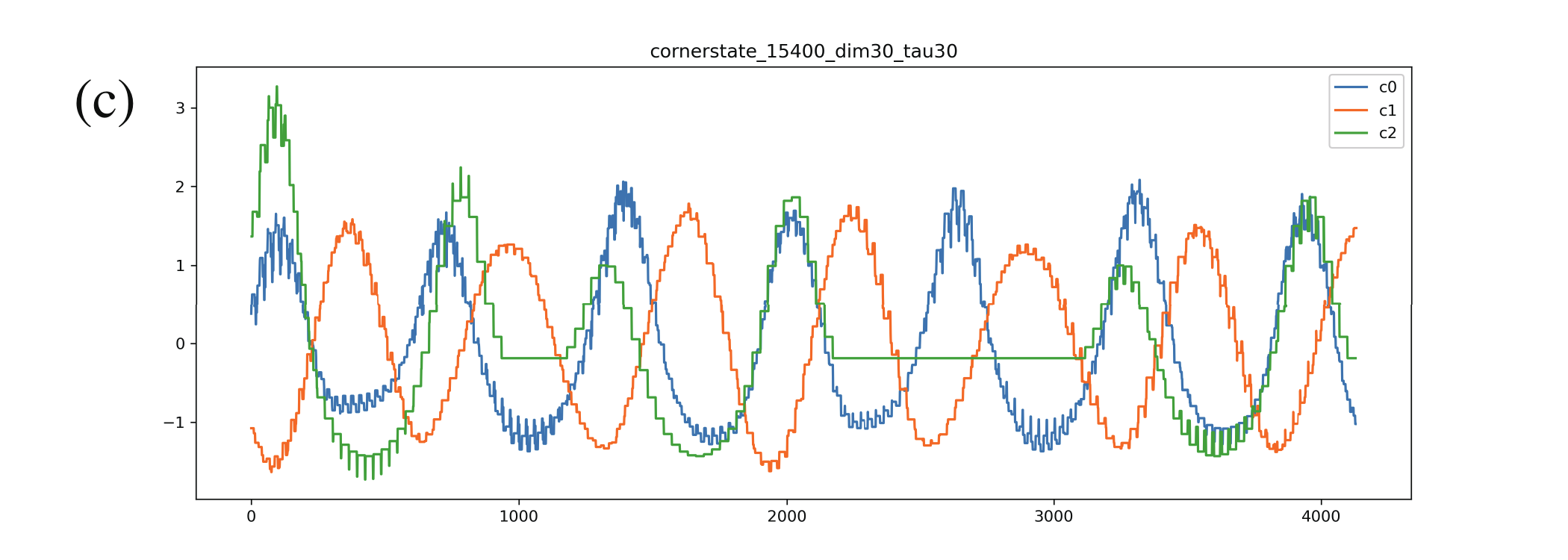}
\includegraphics[bb=0 0 1008 360, width=0.9\textwidth]{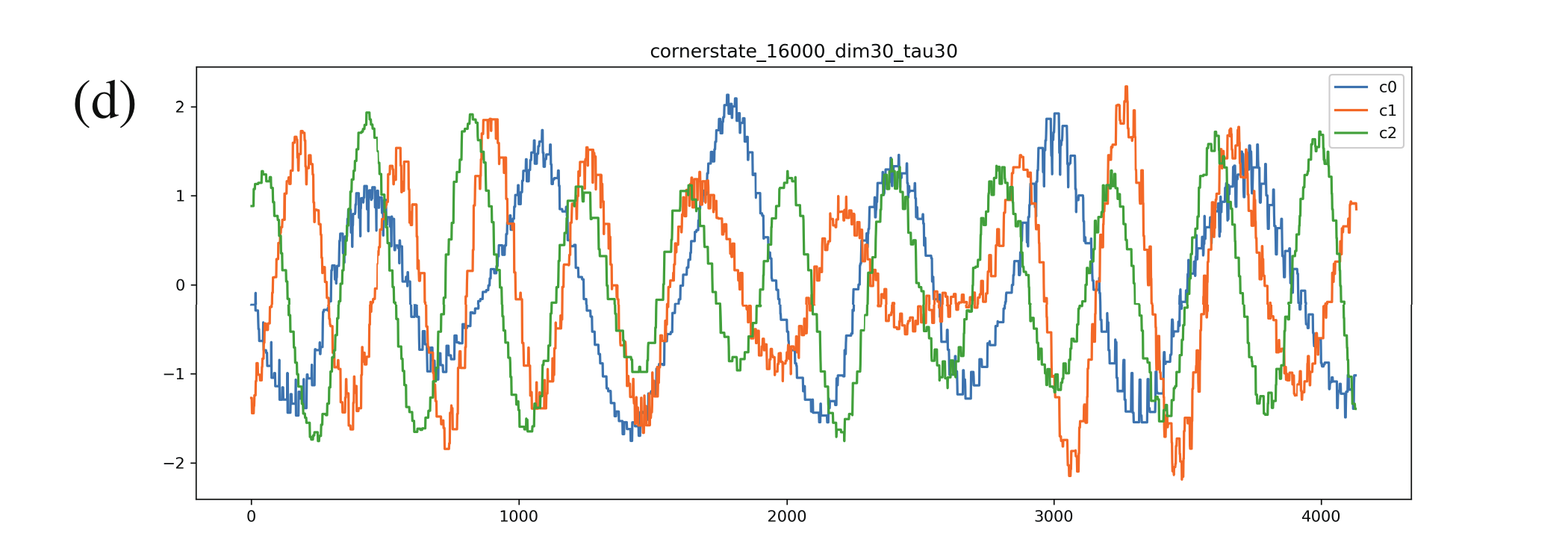}
\end{center}
 \caption{Evolution of corner states at (a) $Re=14000$, (b) $Re=15250$, (c) $Re=15400$, and (d) $Re=16000$.
The plot shows the first principal component of time-delay embedded local structure states $c^0_+, c^1_+$, and $c^2_+$ at the cavity corners. The $x$-axis represents time, while the $y$-axis shows the magnitude of the first principal component.
For Reynolds numbers $Re=14000$ and $Re=15250$, we observe clear periodic behaviour in all corner states.
For $Re=15400$ and $Re=16000$, no apparent periodicity is found, indicating a transition to aperiodic flow.
For $Re=14000$ and $Re=15250$, $c^2_+$ remains constant, suggesting stability in this corner region at lower Reynolds numbers.
 }
\label{fig:corner_state}
\end{figure}

To encode the instantaneous flow state based on COT symbols, the state $v^i_k$ of each corner $i=0,1,2$ at time $t=t_k$ is defined as follows:
$v^i_k=3$ if $b_{++}$ is present inside the parentheses of $c_+^i$; otherwise, it is the smaller of the numbers of $c_+^i$ and $2$. We form a state vector at the $i$-th corner using time-delayed embedding:
\[
u^i_k=(v^i_k, v^i_{k-\tau}, v^i_{k-2\tau}, 
\ldots,  v^i_{k-({\mathcal E}-1)\tau})
\]
with ${\mathcal E}=30$ and $\tau=30$. These values of ${\mathcal E}$ and $\tau$ are determined by maximising the accuracy of the autoregression.
Note that the unit time is $0.01$.
For $Re=14000$ (Figure~\ref{fig:corner_state}(a)) and $Re=15250$ (Figure~\ref{fig:corner_state}(b)), clear periodicity is observed, while no apparent pattern can be seen for  $Re=15400$ (Figure~\ref{fig:corner_state}(c)) and $Re=16000$ (Figure~\ref{fig:corner_state}(d)).

\begin{figure}
\begin{center}
\includegraphics[bb=0 0 818 308, width=0.9\textwidth]{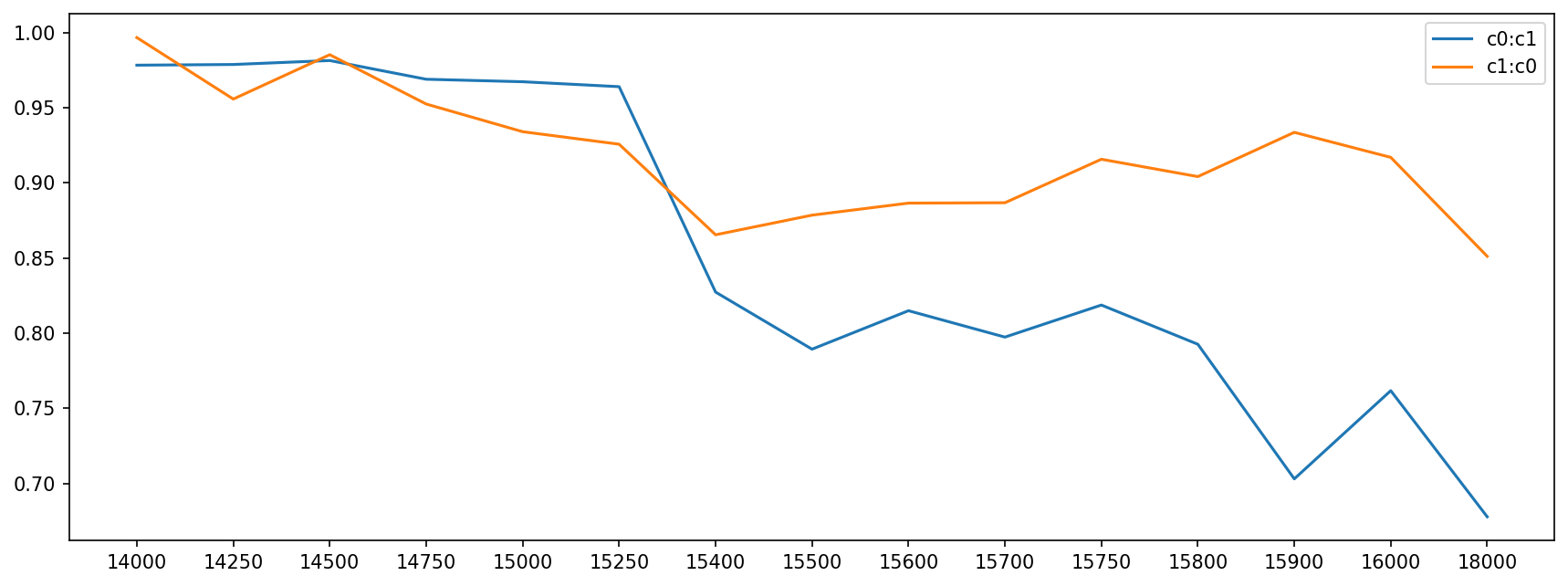}
\end{center}
 \caption{Convergent cross mapping (CCM) analysis of the corner states.
The correlation coefficient between true and predicted states (the $y$-axis) is plotted against Reynolds number (the $x$-axis). 
The blue curve shows prediction accuracy when using $c^0_+$ to predict $c^1_+$, while the orange curve shows the reverse relationship. Higher prediction accuracy (higher correlation between the prediction and the truth) indicate a stronger causal influence from the predicted state to the predictor state, following CCM theory.}
\label{fig:CCM_c0c1}
\end{figure}

To quantify the spatial and temporal causal relation between corners, we quantify how accurately one local structure's state vector predicts another's. 
Since $c^2_+$ remains static for low Reynolds numbers, we focus on the states of $c^0_+$ and $c^1_+$.
Using the time-delayed embedding $u^i_k$ defined earlier, CCM evaluates the precision of the prediction through correlation coefficients, using increasingly larger historical datasets until convergence.
In a deterministic system, the dynamics of the driver's variables is embedded within the effect variables' dynamics. 
The higher prediction accuracy of one variable using the state of another therefore indicates that the predicted variable causally influences the predictor variable.
In \Cref{fig:CCM_c0c1}, we plot two key relationships. 
The orange curve (``c1:c0'') shows the correlation between the predicted and actual state vectors at $c^0_+$, where the predictions use the state vector of $c^1_+$.
A high correlation in this case reveals that the state of $c^0_+$ can be reconstructed from the dynamics of $c^1_+$, indicating that $c^0_+$ causally drives $c^1_+$.
The blue curve (``c0:c1'') represents the reverse relationship.

Up to $Re=15250$, both curves maintain values close to $1$, indicating an indistinguishable bidirectional causal relationship.
This reflects the periodic evolution of the streamline patterns described in Section~\ref{sec:period}, where the topological changes at different corners occur in a coordinated manner.
However, around $Re=15400$, the causation from $c^1_+$ to $c^0_+$ (blue) weakens, while the causation from $c^0_+$ to $c^1_+$ (orange) persists up to $Re=16000$.
This asymmetry suggests that, in the aperiodic regime, topological pattern changes at the top left corner increasingly drive changes at the bottom left corner.
In other words, the transition to aperiodic behaviour is accompanied by a reorganisation of the interaction structure among corner eddies.

These results provide a picture of the transition observed in the COT dynamics.
In the periodic regime, the corner structures evolve in a mutually coupled manner, producing synchronous pattern transitions.
In contrast, once the flow becomes aperiodic, the influence among corners becomes uneven, and the evolution of the system is effectively initiated by structural changes in the top left corner.
Thus, causal analysis based on the COT representation reveals how the hierarchy of interactions among coherent flow structures changes across the transition.
In Appendix \ref{sec:Appendix}, to complement the present observational causal analysis, we further investigate the relation between the two corners through an interventional analysis, which is more closely aligned with control-oriented studies in fluid mechanics.

